\documentclass[11pt]{article}
\usepackage{times}
\usepackage{graphicx,url}
\usepackage[left=1in, right=1in, top=1in, bottom=1in]{geometry}
\usepackage{rotating}

\usepackage{cite}
\usepackage{authblk}
\usepackage{xspace}
\usepackage[table, usenames,dvipsnames]{xcolor}
\usepackage{subfigure, hyperref, graphicx}
\usepackage{amssymb,amsmath,amsthm, booktabs}
\usepackage{algorithm}
\usepackage{color, colortbl}
\usepackage[noend]{algpseudocode}
\usepackage{soul}
\usepackage{placeins}
\usepackage{graphicx}
\usepackage{pgfplots}
\usepackage{amssymb, amsmath, booktabs, times, multirow, verbatim, placeins, dsfont, algorithm, mathtools, color}
\usepackage[normalem]{ulem}
\usepackage[noend]{algpseudocode}
\usepackage[nodisplayskipstretch]{setspace}

\usepackage{xr}

\newenvironment{packed_enum}{
\begin{enumerate}
  \setlength{\itemsep}{1pt}
  \setlength{\parskip}{0pt}
  \setlength{\parsep}{0pt}
}{\end{enumerate}}
\newenvironment{packed_itemize}{
\begin{itemize}
  \setlength{\itemsep}{1pt}
  \setlength{\parskip}{0pt}
  \setlength{\parsep}{0pt}
}{\end{itemize}}

\usepackage[nodisplayskipstretch]{setspace}
\DeclarePairedDelimiter{\ceil}{\lceil}{\rceil}

\newcommand{\OurAlgo}{CoMEt}

\makeatletter
\renewcommand{\@biblabel}[1]{\quad#1.}
\makeatother

\definecolor{offwhite}{RGB}{244,244,244}
\definecolor{offgray}{RGB}{228,228,228}

\begin{document}
%\firstpage{1}
\vspace{0.15in}
\begin{minipage}{0.95\linewidth}
\begin{center}
{\Large
	\textbf{\OurAlgo{}: A Statistical Approach to Identify Combinations of Mutually Exclusive Alterations in Cancer}
} \\
\vspace{0.15in}
{\normalsize
Mark D.M. Leiserson$^{1,2,*}$, Hsin-Ta Wu$^{1,2,*}$, Fabio Vandin$^{1,2,3}$, Benjamin J. Raphael$^{1,2}$\\
}

\vspace{0.25in}

{\small 
\emph{$^1$Department of Computer Science and $^2$Center for Computational Molecular Biology, Brown University, Providence, RI, USA}\\
\emph{$^3$Department of Mathematics and Computer Science, University of Southern Denmark, Odense M, Denmark}\\
\emph{$^{*}$Equal contribution.}\\
	\vspace{0.1in}
	Correspondence:
	\texttt{braphael@brown.edu}
}
\end{center}
\end{minipage}
\vspace{0.25in}

\begin{abstract} % abstract
Cancer is a heterogeneous disease with different combinations of genetic and epigenetic alterations driving the development of cancer in different individuals.  While these alterations are believed to converge on genes in key cellular signaling and regulatory pathways, our knowledge of these pathways remains incomplete, making it difficult to identify driver alterations by their recurrence across genes or known pathways. We introduce \underline{C}ombinations \underline{o}f \underline{M}utually \underline{E}xclusive  Al\underline{t}erations (\OurAlgo{}), an algorithm to identify \emph{combinations} of alterations \textit{de novo}, without \emph{any} prior biological knowledge (e.g. pathways or protein interactions). \OurAlgo{} searches for combinations of mutations that exhibit mutual exclusivity, a pattern expected for mutations in pathways.

\OurAlgo{} has several important feature that distinguish it from existing approaches to analyze mutual exclusivity among alterations.  These include: an exact statistical test for mutual exclusivity that is more sensitive in detecting combinations containing rare alterations;  simultaneous identification of collections of one or more combinations of mutually exclusive alterations; simultaneous analysis of subtype-specific mutations; and summarization over an ensemble of collections of mutually exclusive alterations. These features enable \OurAlgo{} to robustly identify alterations affecting multiple pathways, or hallmarks of cancer. We show that \OurAlgo{} outperforms existing approaches on simulated and real data.  Application of  \OurAlgo{} to hundreds of samples from 4 different cancer types from TCGA reveals multiple mutually exclusive sets within each cancer type.  Many of these overlap known pathways, but others reveal novel putative cancer genes.
\end{abstract}

\section{Introduction}
A major goal of large-scale cancer genomics projects such as The Cancer Genome Atlas (TCGA) \cite{TCGAGBM2008,TCGAGBM2013,TCGAAML,TCGABRCA,TCGAPanCan,TCGASTAD}, the International Cancer Genome Consortium (ICGC) \cite{ICGC,ICGCDataPortal}, and others is to identify the genetic and epigenetic alterations that drive cancer development. These projects have generated whole-genome/exome sequencing data measuring the somatic mutations in thousands of tumors in dozens of cancer types.  Interpreting this data requires one to distinguish the \emph{driver} mutations that play a role in cancer development and progression from  \emph{passenger} mutations that have no consequence for cancer. Identifying driver mutations directly from sequencing data is a significant challenge since individuals with the same cancer type typically exhibit different combinations of driver mutations \cite{Lawrence2013,Salk2010}. This mutational heterogeneity arises because driver mutations target genes in pathways -- collections of interacting genes that perform a biological function (e.g. signaling or regulation) -- such that each pathway can be perturbed in numerous ways~\cite{Vogelstein2013}.

The observed mutational heterogeneity in cancer has motivated the development of methods to examine \emph{combinations} of mutations, including methods that examine known pathways or networks (reviewed in \cite{Raphael2014, Ding2013}).  However, most pathway databases and interaction networks are incomplete, lack tissue-specificity, and do not accurately represent the biology of a particular cancer cell. Thus, \emph{de novo} methods for examining combinations of mutations are of particular interest as they require no prior biological knowledge and enable the discovery of novel combinations.  Unfortunately, the number of possible combinations is too large to test exhaustively and achieve statistically significant results.  Current \emph{de novo} approaches to identify putative combinations of mutations use the observation that mutations in the same pathway are often mutually exclusive \cite{Yeang2008}. This observation follows from the observation that there are relatively few driver mutations in a tumor sample, and these are distributed over multiple pathways/hallmarks of cancer \cite{Hanahan2011}.

In 2011, three algorithms for identifying sets of genes with mutually exclusive mutations were introduced simultaneously:  the De Novo Driver Exclusivity (Dendrix)~\cite{Dendrix}, Recurrent Mutually Exclusive aberrations (RME)~\cite{RME}, and Mutual Exclusivity Modules (MEMo)~\cite{MEMo} algorithms.  Dendrix and RME are both \emph{de novo} algorithms for identifying gene sets with mutually exclusive mutations, while MEMo examines mutual exclusivity on a protein-protein interaction network. The Dendrix algorithm identifies sets $M$ of $k$ genes with high coverage (many samples have a mutation in the set) and approximate exclusivity (few samples have a mutation in more than one gene in the set).  Dendrix combines these two criteria into a weight $W(M)$, which is equal to the coverage of $M$ minus the coverage overlap (co-occurring mutations) of $M$.  Finding the set of maximum weight is an NP-hard problem~\cite{Dendrix}.  Dendrix uses a Markov chain Monte Carlo (MCMC) algorithm to sample high weight gene sets; more recently other optimization methods have been used to find high weight sets \cite{Zhao2012,Li2014}.   Leiserson \textit{et al}.~\cite{MultiDendrix} introduced the Multi-Dendrix algorithm to identify multiple mutually exclusive gene sets simultaneously using an integer linear program. In contrast, RME defines the exclusivity weight as the percentage of covered samples that contain exactly one mutation within a gene set, and uses an online-learning linear threshold algorithm to identify groups of genes with high pairwise exclusivity.  However, both the RME and MEMo algorithms were shown not to scale to reasonable-sized datasets~\cite{MultiDendrix}, requiring extensive filtering of input data \cite{RME,Ciriello2013}.

\begin{figure*}[h!]
\centering
\includegraphics[width=\textwidth]{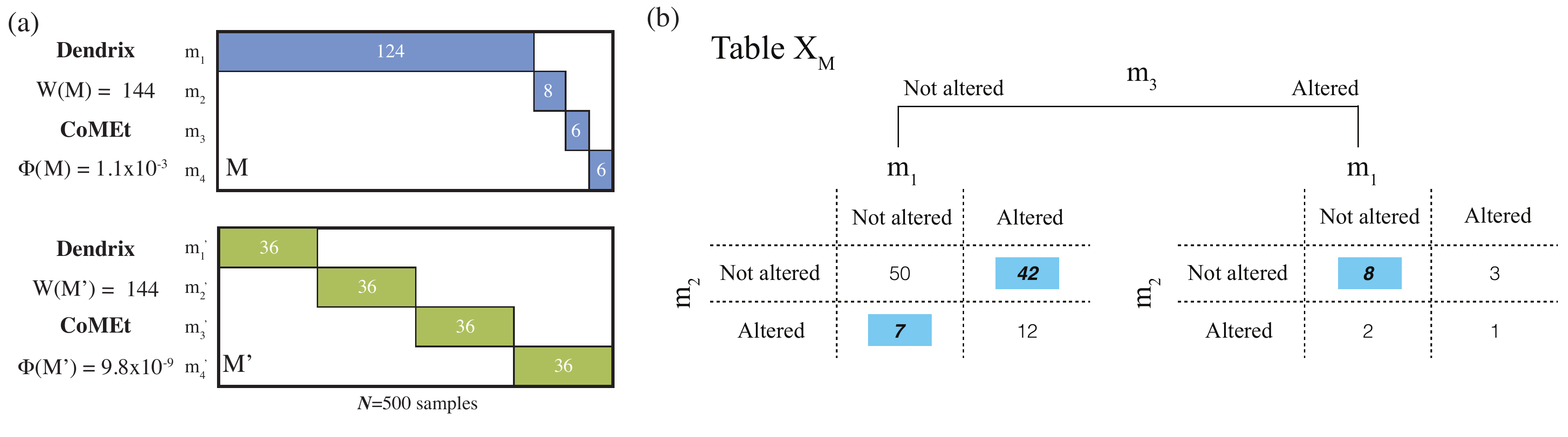}
\caption{(a) \textbf{Alteration matrices illustrating differences between the combinatorial weight function $W(M)$ introduced in Dendrix and the probabilistic score $\Phi(M)$ used in \OurAlgo{}.}  Both matrices contain 4 mutually exclusive alterations whose alteration frequencies are indicated inside each bar. The samples without alterations are not shown in either matrix. Since both sets are exclusive and have the same total alteration frequency, the Dendrix weight function does not distinguish between these sets.  Sets like $M$ (left) are common in cancer genome studies which often have a small number of recurrently mutated genes and a long tail of rarely mutated genes.  The score used in \OurAlgo{} conditions on the observed frequencies of each alteration, giving more significance to the set $M'$. (b) \textbf{An example of $2 \times 2 \times 2$ contingency table $\mathbf{X}_M$ for the set $M= \{m_1, m_2, m_3\}$, illustrating how samples are cross-classified into exclusive, co-occurring, or absent for each alteration.} The test statistic $\phi(M)$ used by \OurAlgo{} is the sum of the highlighted exclusive cells.}
\label{fig:small_model}
\end{figure*}

One limitation of the combinatorial weight function used in Dendrix and subsequent algorithms is that genes with high mutation frequencies (high coverage) can dominate the mutual exclusivity signal, thus biasing the algorithms towards identifying gene sets where the majority of the coverage comes from one gene (Figure~\ref{fig:small_model}(a)).  These observations motivated the development of probabilistic models of mutual exclusivity.  These include the Dendrix++ algorithm (an early version of the approach that we present in this paper) and the muex algorithm~\cite{muex}.  Dendrix++ uses a statistical score and was used in TCGA acute myeloid leukemia study \cite{TCGAAML}.  The muex algorithm~\cite{muex} uses a generative model of mutual exclusivity and a likelihood ratio test to identify mutually exclusive sets.  We find that muex remains sensitive to the presence of high frequency mutations (See Section \ref{sec:comparison-real-data}).  Moreover, both of these approaches exhaustively enumerate gene sets to find those with high score, limiting their applicability to larger datasets.  In addition, they do not identify multiple gene sets simultaneously, a feature that has proved useful with the Dendrix weight \cite{MultiDendrix}.  Finally, no current method identifies overlapping gene sets\footnote{We note that while Multi-Dendrix~\cite{MultiDendrix} allows for searching for overlapping gene sets, this option was never explored.}, although cancer genes have been shown to participate in multiple pathways \cite{TCGAGBM2008}, or addresses the problem of cancer subtype-specific mutations which can confound the mutual exclusivity signal.  

We introduce the \underline{C}ombinations \underline{o}f \underline{M}utually \underline{E}xclusive  Al\underline{t}erations (\OurAlgo{}) algorithm to address the limitations outlined above.  \OurAlgo{} includes the following contributions.
% the following key features.
\begin{packed_enum}
\item We develop an exact statistical test for mutual exclusivity \emph{conditional} on the observed frequency of each alteration. This approach is less biased towards high frequency alterations, and enables the discovery of combinations of lower frequency alterations. We derive a novel tail enumeration procedure to compute the exact test, as well as a binomial approximation.
\item \OurAlgo{} simultaneously identifies collections consisting of \emph{multiple} combinations of mutually exclusive alterations, and samples from such collections using an MCMC algorithm.  We summarize the resulting distribution by computing the marginal probability of pairs of alterations in the same sets. This enables \OurAlgo{} to identify sets of any size, including overlapping sets of alterations, without testing many parameter settings.
\item Given prior knowledge of cancer-types/subtypes, \OurAlgo{} analyzes alterations and subtypes simultaneously, allowing the discovery of mutually exclusive alterations across cancer types, while avoiding the identification of spurious mutually exclusive sets of (sub)type-specfic mutations.
\end{packed_enum}

We demonstrate that \OurAlgo{} outperforms earlier approaches on simulated and real cancer data.  We apply \OurAlgo{} to acute myeloid leukemia (AML), glioblastoma (GBM), gastric (STAD), and breast cancer (BRCA) data from TCGA.  In each cancer type, we identify combinations of mutated genes that overlap known cancer pathways and and also contain potentially novel cancer genes including \textit{IL7R} and the EphB receptor \textit{EPHB3} in STAD, and the scavenger receptor \textit{SRCRB4D} in GBM. On the gastric and breast cancer data, we demonstrate how \OurAlgo{} simultaneously identifies mutual exclusivity resulting from pathways and from subtype-specific mutations.  \OurAlgo{} is available at \url{http://compbio.cs.brown.edu/software/comet}.

\section{Methods} \label{sec:method_material}

\begin{figure*}[h!]
\begin{center}
  \includegraphics[width=0.95\textwidth]{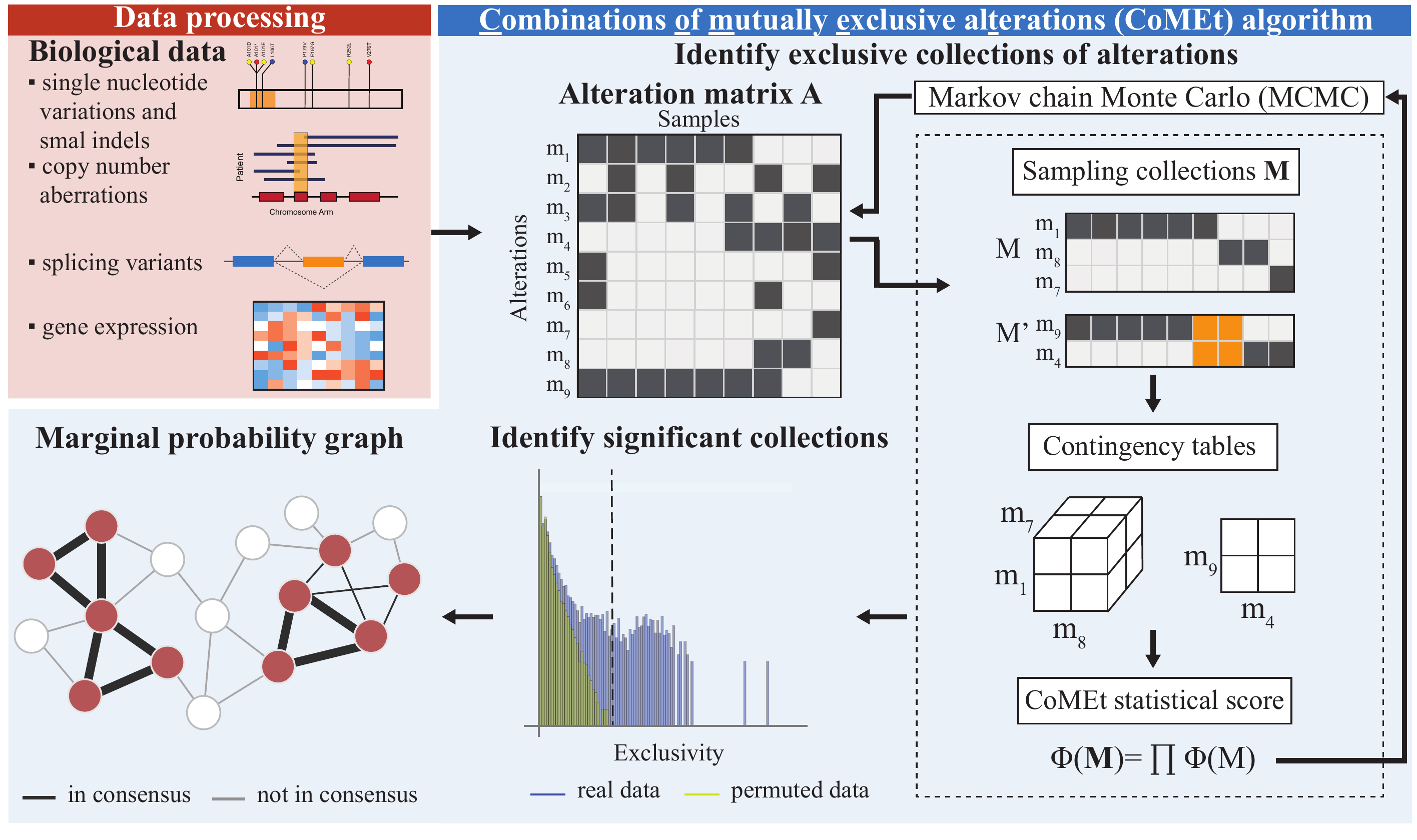}
  \caption{\textbf{Overview of the \OurAlgo{} algorithm.} First, we transform alteration data from different measurements into a binary  alteration matrix $A$. Second, for fixed values of $k$ and $t$ we use a Markov Chain Monte Carlo (MCMC) algorithm to sample collections $\mathbf{M}$ in proportion to the weight $\Phi(\mathbf{M})^{-1}$.  Here we show a collection containing sets $\mathbf{M}$ and $\mathbf{M}'$ with three and two alterations, respectively.  We identify all collections whose weight exceeds the maximum observed in randomly permuted datasets.  We summarize the alterations in these significant collections with a \emph{marginal probability graph}, whose edge weights indicate the fraction of significant collections with the corresponding pair of alterations.
 }
  \label{fig:overview}
\end{center}
\end{figure*}

\subsection{Overview of the \OurAlgo{} Algorithm}
We consider that a set $\mathcal{E}$ of $m$ \emph{alterations} have been measured in $n$ samples.  
An alteration may be the somatic mutation of a particular gene,  a specific single nucleotide mutation (e.g. V600E mutations in the \emph{BRAF} gene), an epigenetic change such as hypermethylation of a promoter, or a variety of other changes.  
We assume that alterations are binary, such that alterations are either present or absent in each sample.
We represent the set of measured alterations with an $m \times n$ binary alteration matrix $A = \left[a_{ij} \right]$, where $a_{ij} = 1$ if alteration $i$ occurs in sample $j$, and $a_{ij} = 0$ otherwise. 
Our goal is to identify \emph{one or more} sets $M_1, M_2, \dots, M_t$ where the alterations in each $M_i$ are surprisingly mutually exclusive across the $n$ samples.  We introduce the \OurAlgo{} algorithm for this purpose.

\OurAlgo{} uses a novel statistical score based on an exact test for mutual exclusivity. Figure~\ref{fig:small_model} motivates the development of the new score, showing two sets $M$ and $M'$, each with four alterations.  The alterations in both sets are perfectly exclusive (no sample has more than one alteration), and the total number of altered samples is the same.  The Dendrix weight function $W(M)$ introduced in \cite{Dendrix} (and used in later publications \cite{Zhao2012,Li2014,MultiDendrix}) is defined as the \emph{coverage}, the number of samples with at least one mutation in $M$, minus the \emph{coverage overlap}, the number of samples with more than one mutation in $M$.  In this case, $W(M) = W(M')$.  However, given the frequencies of each alteration, we are more surprised to observe mutual exclusivity among alterations in the set $M'$, which are each altered in $7\%$ of samples, than we are to observe mutual exclusivity among the alterations in set $M$, where a single alteration has very high frequency (25\%) and three alterations have relatively low frequency ($<2\%$).  Sets like $M$ are common in many cancer datasets where highly recurrent alterations (e.g. mutations in \textit{TP53} or amplification of \textit{EGFR}) occur and can be combined with low frequency, spurious alterations.  This problem was noted in \cite{muex}, but the probabilistic model introduced therein seems to overcorrect for this effect, missing important combinations of alterations (See Section \ref{sec:comparison-real-data}).

We derive a score $\Phi(M)$ for a set $M$ of $k$ alterations using an exact test of mutual exclusivity.
Specifically, we examine a $2 \times 2 \times \dots \times 2 = 2^k$ contingency table $\mathbf{X}_M$ (Figure~\ref{fig:small_model}(b)) whose entries indicate the number of samples where each combination of alterations occur.  For example,  the entry $x_{(24)}$ of $\mathbf{X}_M$ equals the number of samples where the second and fourth alterations in $M$ occur, but the first and third alterations do not occur.  
%\ben{[Note that the notation $x_{(24)}$ might not be clear to readers at this point in text.  Not sure if there is an easy solution to this.]}
The score $\Phi(M)$ is the $P$-value of the observed mutual exclusivity in the table $\mathbf{X}_M$, where the margins of the table (determined by the number of samples where each alteration occurs) is fixed.  That is the score $\Phi(M)$ is \emph{conditional} on the observed  frequencies of alterations in $M$. This statistical score reduces the effect of the most frequent alterations have an unduly large contribution to the score.  

\OurAlgo{} scores a collection $\mathbf{M} = (M_1, \dots, M_t)$ of $t$ alteration sets by taking the product of the scores of each set $M_i$:
\begin{equation}
\Phi(\mathbf{M}) = \prod_{i=1}^t \Phi(M_i).
\end{equation}
This score follows from the null hypothesis that exclusivity is independent across sets.

Since the number of possible collections of alteration sets grows exponentially with the number of alterations, it is typically impossible to enumerate and compute the weight of all alteration sets.  We use a Markov chain Monte Carlo (MCMC) algorithm to sample collections $\mathbf{M}$ of $t$ alteration sets, where each collection is sampled with proportion to its weight $\Phi(\mathbf{M})^{-1}$. We summarize this distribution by computing the marginal probability $p(e, e')$ for each pair of alterations in $A$.   We summarize these probabilities using the \emph{marginal probability graph}, a complete, undirected weighted graph $G=(V, E)$ where $V=\mathcal{E}$ and where each edge $e \in E$ connects a pair of vertices $u, v$ with weight $p(u,v)$.   We identify the most exclusive alteration sets by first removing all edges from the graph weight below a threshold $\delta$.  The output of \OurAlgo{} is $C(\delta)$, the connected components in the resulting graph, which we call \textit{modules}.
The summarization via the marginal probability graph allows \OurAlgo{} to output collections of alteration sets different in number and size than specified by the input parameters.  
%\subsection{\OurAlgo{} Algorithm}
%We consider a set of $m$ \emph{alterations} measured in $n$ samples. An \emph{alteration} can be a variety of different genomic, transcriptomic, or epigenomic changes measured in a cancer sample; e.g. the somatic mutation of gene,  a mutation in a particular amino acid residue (e.g. V600E mutations in the \textit{BRAF} gene that are common in colorectal and other cancers \cite{Davies2002}, or an epigenetic change such as hypermethylation of a promoter). We assume that alterations are binary: in each sample, an alteration either occurs or does not occur. We represent the status of $m$ measured alterations in $n$ samples with an $m \times n$ binary alteration matrix $A = \left[a_{ij} \right]$, where $a_{ij} = 1$ if  alteration $i$ occurs in sample $j$, and $a_{ij} = 0$ otherwise. We define a set of $k$ measured alterations as a $n\times k$ submatrix $M$. Our goal is to identify a collection $\mathbf{M} = (M_1, M_2, \dots, M_t)$ of \emph{one or more} sets of mutually exclusive alterations across the $n$ samples.   We introduce the \underline{C}ollections \underline{o}f \underline{M}utually \underline{E}xclusive  Al\underline{t}erations (\OurAlgo{}) algorithm for this purpose (See Figure~\ref{fig:overview}).

\subsection{Scoring mutual exclusivity}

We first describe a statistical score $\Phi(M)$ for a tuple $M = (m_1, \dots, m_k)$ of alterations.  The score measures the surprise of the observed exclusivity of these alterations  \emph{conditional} on the rate of occurrence of each alteration.  Since these rates are generally unknown (e.g. the background mutation rate for single nucleotide mutations varies greatly across genes and samples \cite{Lawrence2014}), we use the \emph{exact distribution} obtained from the observed data as the null distribution.  Under this distribution, the status of the $k$ alterations in $n$ samples is described by selecting uniformly a $k \times m$ binary alteration matrix $B$ with the constraint that the number of 1's in row $i$ of $B$ equals the number of 1's in row $m_i$ of the alteration matrix $A$.  This distribution is equivalent to the sampling distribution on $2 \times 2 \times \dots \times 2 = 2^k$ contingency tables under the hypergeometric distribution, where dimension $i$ of the table gives the cross-classification of the number of samples where alteration $i$ occurs or not.  For example, three alterations are described by a $2 \times 2 \times 2$ table with margins equal to the frequency of each alteration (Figure~\ref{fig:small_model}(b)).

We introduce notation to describe the statistical test. Given a set $M$ of alterations, let $x^{+}_{(j)}$ be the number of samples where alteration $m_j$ occurs. It follows that $n - x^{+}_{(j)}$ is the number of samples where $m_j$ does not occur.  Similarly, for $\mathbf{v} \subseteq [k] = \{1, \dots, k\}$, let $x_\mathbf{v}$ denote the number of samples where alterations only occur in $m_{\mathbf{v}}$. The values $x_{\mathbf{v}}$ for all $v \subseteq [k]$ give the entries of a $2^k$ contingency table $\mathbf{X}_M$ with fixed margins $\mathbf{x^+} = (x^{+}_{(1)}, \dots, x^{+}_{(k)})$. Thus, the probability of observing a $2^k$ contingency table $\mathbf{X}_M$ with fixed margins $\mathbf{x^+}$ and whose sum of entries equals $n$ follows the  multivariate hypergeometric distribution
\begin{equation}
\label{eqn:hypergeom}
p_{\mathbf{X}_M} = \Pr(\mathbf{X}_M|  \mathbf{x^+}, k, n) = \frac{ \prod_{j=1}^k x^{+}_{(j)}! \left( n -x^{+}_{(j)} \right)! }{ (n!)^{k-1} \prod_{\mathbf{v} \subseteq [k]} x_{\mathbf{v}}!} .
\end{equation}

To characterize the mutual exclusivity of alterations in a contingency table, we define the test statistic as the sum of the entries in the contingency table where \emph{exactly} one alteration occurs, i.e. $T(\mathbf{X}_M) =\sum_{j=1}^k x_{\{j\}}$, where $x_{\{j\}}$ is the number of samples where alterations occur only in $m_j$. We compute a $P$-value for the observed value $T(\mathbf{X}_M)$ of the test statistic as the tail probability of observing tables with the same margins whose exclusivity is at least as large as observed:
\begin{equation}
\label{eqn:exact_tail}
\Pr(T \ge T(\mathbf{X}_M) | \mathbf{x^+}, k, n ) = \sum_{ \substack{\mathbf{Y} \in \mathcal{T}(\mathbf{x^+}): \\ T(\mathbf{Y}) \ge T(\mathbf{X}_M)}} \Pr(\mathbf{Y} |  \mathbf{x^+}, k, n ),
\end{equation}

where $\mathcal{T}(\mathbf{x^+})$ is the set of $2^k$ contingency tables with margins $\mathbf{x}^+$.   Note that for $k = 2$, the test statistic $T(\mathbf{X}_M)$ is equivalent to a one-sided Fisher's exact test.   $2 \times 2$ contingency tables have only one degree of freedom, and thus there are essentially only two ways in which the corresponding pair of random variables can be non-independent: having too many co-occurrences or too much exclusivity (Figure~\ref{fig:small_model}(b)). However, $2^k$ tables have $2^k - k -1$ degrees of freedom and there are many ways in which the corresponding random variables can be non-independent.  The $T(\mathbf{X}_M)$ test statistic measures whether the alterations are surprisingly \emph{mutually exclusive}, rather than non-independent in some other way. 

We define the score $\Phi(M)$ using the mid $P$-value \cite{Lancaster1961}, which is the the average of the probability of observing a value at least as extreme as the observed value and observing a value more extreme than observed:
\begin{equation}
\label{eqn:mid_pval_weight}
\Phi(M)  =  \frac{1}{2} (Pr(T \ge T(\mathbf{X}_M) | \mathbf{x^+}, k, n) + Pr(T > T(\mathbf{X}_M) | \mathbf{x^+}, k, n) ).
\end{equation}
We use the mid $P$-value because the tail probability from exact tests is typically overly conservative, due to the discreteness of the exact distribution \cite{Lancaster1961}. Finally, since cancer is driven by mutations in multiple pathways \cite{Hanahan2011}, we define a score  $\Phi(\mathbf{M})$ for a collection $\mathbf{M}=(M_1, M_2, \dots, M_t)$ of $t$ gene sets as $\Phi(\mathbf{M}) = \prod_{i=1}^t \Phi(M_i)$. The product results from our assumption that under the null hypothesis mutations in different sets $M_i$ are independent.

\subsection{Computing the mutual exclusivity score $\Phi(M)$}
To compute the mutual exclusivity score $\Phi(M)$, one must compute \eqref{eqn:exact_tail}.  This requires computing the probability of all tables $\mathbf{Y}$ with the same margins as $\mathbf{X}_M$ and with exclusivity statistic $T(\mathbf{Y})$ at least as large as the observed value $T(\mathbf{X}_M)$.  Unfortunately, no algorithm is known to enumerate such tables.  In general the problem of counting contingency tables with fixed margins are \#P-complete \cite{Dyer1997}, and thus it is unlikely they can be enumerated efficiently.  Several methods have been proposed to solve the problem of counting contingency tables, including using the network algorithm \cite{mehta1983network, requena2006major} for Fisher's exact test in $r \times c$ contingency tables, or extensions to consider joint effect of two contingency tables (i.e. $2\times r\times c$) \cite{de2005additive}. Branch and bound heuristics have also been used in some specialized cases \cite{bejerano2004efficient}. However, these approaches still consider at most three dimensional contingency tables, and the problem of enumerating $2^k$ tables does not seem to have been considered.
 Even for small $k$ the enumeration problem is intractable: the number of $2^k$ tables with fixed margins grows exponentially in $k$. \cite{Zelterman1995} presented an exhaustive algorithms to enumerate all $2^3$ and $2^4$ contingency tables with fixed margins, demonstrating for example that for $n=36$, there are $>100$ million $2^4$ tables.
Randomized and approximate counting methods for contingency tables have been developed (e.g. \cite{Barvinok2010,Miller2013} and references therein), although these generally do not provide a rigorous guarantee on the error in the approximation.

We derive a novel \emph{tail enumeration} algorithm to efficiently compute the tail probability in equation~\eqref{eqn:exact_tail} for tables with high values of the exclusivity statistic $T$. 
The motivation for our approach is that the sets $M$ of interest will have extremely high values of $T(\mathbf{X}_M)$, near the maximum possible value.  For example, in the degenerate case of perfect exclusivity (no sample with more than one alteration in $M$) there are no more extreme tables to enumerate, and the algorithm needs only to evaluate the hypergeometric probability of equation~\eqref{eqn:hypergeom} for this single table. Thus, if we enumerate tables starting from the highest possible values for $T$, we can obtain highly accurate $P$-values for the most interesting cases.  Furthermore, we can stop the enumeration procedure when the $P$-value becomes sufficiently large and use approximations for these larger $P$-values (See below).

Algorithm~\ref{alg:exact_general} is the tail enumeration strategy to enumerate contingency tables in approximate order from most to least exclusive. Briefly, let $\mathbf{C}=( \mathbf{v} \subseteq [k] :  |\mathbf{v}| \ge 2 )$ be the vector of co-occurring (not exclusive) cells. The basic strategy employed by Algorithm~\ref{alg:exact_general} is to generate a table $\mathbf{Y}$ that is more exclusive than $\mathbf{X}_M$ (i.e. $T(\mathbf{Y}) > T(\mathbf{X}_M)$ by iterating through the possible values of each cell in $\mathbf{C}$, using the following facts:
\begin{packed_itemize}
\item When all values in $\mathbf{C}$ are fixed, the other values in the contingency table are uniquely determined (See Procedure \textsc{CompleteContTbl} in Algorithm~\ref{alg:exact_general}).
\item We can set and update exact upper and lower bounds for each cell in $\mathbf{C}$. The values of each cell are bounded by two values (lines 10-11 in \textsc{TailEnumeration}): the first is how many more co-occurrences are allowed in the current table ($T_{REM}$) before $\mathbf{Y}$ is less exclusive than $\mathbf{X}_M$; the second is given by the constrained marginal ($MarRem$) for that variable in $\mathbf{X}_M$.
\end{packed_itemize}
We find that Algorithm~\ref{alg:exact_general} performs well on real data, evaluating the test statistic $T(\mathbf{X}_M)$ in a few seconds for sets with $k \le 7$ that have a small number of co-occurrences.

\begin{algorithm}[ht]
\renewcommand{\algorithmicrequire}{\textbf{Input:}}
\renewcommand{\algorithmicensure}{\textbf{Output:}}
\caption{Tail enumeration for any $k > 1$}
\label{alg:exact_general}
\begin{algorithmic}[1]
 \Require $2^k$ contingency table $\mathbf{X}$.
 \Ensure Set $\mathcal{S}$ of contingency tables at least as exclusive as $\mathbf{X}$: $\mathcal{S} = \mathbf{Y} \in \mathcal{T}(x^+) : T(\mathbf{Y})\ge T(\mathbf{X})$.
 
 \State $\mathcal{S} \gets \{  \}$
 \State $N \gets 2^k$
 \State $\mathbf{C} \gets $\Call{Sorted}{$\{\mathbf{v} \subseteq [k] : |\mathbf{v}| \ge 2\}$}\Comment{Sorted descending vector of co-occurring cells}
 \State $y_\mathbf{v} \gets 0, \forall \mathbf{v} \subseteq [k]$
 \State $T_{\max} \gets \sum_{i=1}^k x^+_{(i)}$\Comment{Sum of alteration frequencies}
 \State \Call{TailEnumeration}{$\mathbf{Y}, \mathbf{C}, T_{\max} - T(\mathbf{X})$}
 
 \Procedure{TailEnumeration}{$\mathbf{Y}, \mathbf{C}, T_{REM}$} \Comment{$T_{REM}$: count of allowed co-occurrences remaining}
	\State $\mathbf{v} \gets $ \Call{Head}{$\mathbf{C}$}
	\If {$\mathbf{v} \neq $ \textsc{Null}}
		\State $MarRem \gets \underset{i \in \mathbf{v}}{\min \text{ }}{\{ y^+_i \}}$
		          \Comment{Minimum margin remaining}
		\For{$(i \gets L,\dots,\min\{MarRem, \lfloor \frac{T_{REM}}{|\mathbf{v}|}\rfloor\}) $}
			\State $\mathbf{Y}' \gets \Call{Copy}{\mathbf{Y}}$
			\State $y'_\mathbf{v} \gets i$\Comment{Set value of cell $\mathbf{v}$ of $\mathbf{Y}'$ to $i$}
			\State \Call{TailEnumeration}{$\mathbf{Y}', $\textsc{Tail}$(\mathbf{C}), T_{REM} - |\mathbf{v}| \times i$}
		\EndFor
	\Else \Comment{If all ``co-occurring'' cells have been set}
		\State $\mathcal{S} = \mathcal{S} \cup \{ \Call{CompleteContTbl}{\mathbf{Y}} \}$
	\EndIf
 \EndProcedure
  \Procedure{CompleteContTbl}{$\mathbf{Y}$} \Comment{Fill in remainder of contingency table $x'$}
	\For {$\mathbf{v} \subseteq [k] : |\mathbf{v}| = 1$}\Comment{Iterate over exclusive cells}
		\State $y_\mathbf{v} \gets x^+_\mathbf{v} - y^+_\mathbf{v}$
	\EndFor 
	\State $\mathbf{Y}_{(0, 0, \dots, 0)} \gets n - \sum_{y \in \mathbf{Y}} y$ \Comment{Fill in cell with no alterations}
	\State \Return $\mathbf{Y}$
 \EndProcedure
 \label{appendix:exact_test}
\end{algorithmic}
\end{algorithm}

\paragraph{Binomial approximation.}
We can approximate the distribution of the exclusivity statistic using the binomial distribution, which is a  well-known approximation of the hypergeometric distribution. Under the null hypothesis that alterations occur independently in the samples, let $p_e = \sum_{j=1}^k \frac{x_{(j)}}{n}$ be the probability of an exclusive alteration; i.e. a sample contains exactly one alteration in $M$. Given a set $M$ of alterations $M$, then the probability of observing $T(\mathbf{X}_M)$ or more exclusive alterations in $n$ samples is given by the binomial tail probability $1 - \sum_{i=0}^{T(\mathbf{X})-1} {n \choose i} p_e^i (1-p_e)^{n-i}$.

We find that the binomial provides a good approximation of the exact test $P$-value for sets $M$ with a large number of co-occurring mutations, and consequently a higher $P$-value (See Figure~\ref{fig:approx_binom_exact_vs_cooccurring}).  Conveniently, these are precisely the cases where the tail enumeration algorithm is slow.

\paragraph{Permutation approximation.}
Another approximation to the exact test is obtained using a permutation test.  We sample $L$ tables with fixed margins uniformly from the space of all tables and compute the proportion of such tables whose exclusivity value $T$ exceeds the observed value $T(\mathbf{X}_M)$.  Of course, sampling uniformly from the set of tables with fixed margins is not straightforward.  We use an MCMC approach as described in \cite{MEMo}, although we do not fix the number of alterations per sample. Interestingly, while the MEMo algorithm \cite{MEMo} uses a permutation test, the test statistic is the coverage $\Gamma(M)$, rather than the exclusivity $T(M)$ used in \OurAlgo{}.  While these are equivalent when $k = 2$ (since there is only one degree of freedom), they produce different results for $k > 2$.  See further discussion in Section \ref{sec:comparison-to-memo}.

In our implementation, we use the exact test, binomial approximation, or permutation approximation to compute $\Phi(M)$ according to the following procedure.  First, we calculate the $P$-value from the binomial approximation and compute the number of co-occurring  alterations in $M$.  If the number of co-occurring  alterations is higher than a fixed threshold $\kappa$ or the binomial $P$-value is larger than a fixed value $\psi$, we set $\Phi(M)$ to be the binomial $P$-value.  Otherwise, we perform the tail enumeration procedure to compute the exact test $P$-value, stopping the enumeration if the accumulated tail probability becomes larger than a threshold $\epsilon$. If we stop, then we compute the permutation approximation with $\ceil{\frac{1}{\epsilon}}$ samples, such that we expect to sample at least one table with $T > T(\mathbf{X}_M)$. This procedure focuses the time to perform tail enumeration  in those cases where high accuracy is needed for small $P$-values.

\subsection{Sampling collections of mutually exclusive alterations with MCMC}
\label{sec:mcmc}
Our goal is to identify a collection $\mathbf{M}$ of $t$ alteration sets with low (highly significant) values of $\Phi(\mathbf{M})$.  Since is typically not possible to enumerate all such collections (except for test datasets with small $m$, $n$, $t$, and $k$), we derive a Markov Chain Monte Carlo (MCMC) approach to sample from the space of possible collections. We use the Metropolis-Hastings algorithm \cite{Metropolis1953,Hastings1970} to derive a Markov chain Monte Carlo (MCMC) algorithm to sample collections $\mathbf{M}$ in proportion to the weight $\Phi(\mathbf{M})^{-1}$ (See Supplementary Section \ref{sec:mcmc-alg}). 

\subsection{Marginal probability graph} 
\label{sec:marginal-prob-graph}
We now present a method to extract a collection of highly exclusive alteration sets (\emph{with no prescribed size}) from the posterior distribution obtained from the MCMC algorithm. Typically, there are multiple collections with significant scores.  This might be for interesting reasons such as different sets of alterations with similar scores or alterations that appear in multiple mutually exclusive sets.  However, the reason might also be suboptimal parameter selection; e.g. there may be a significant set of $k=3$ alterations in the data, but running the algorithm with $k=4$ will return many sets with the same three genes and a fourth ``spurious" gene.  To distinguish such cases, we summarize the posterior distribution on collections using a \emph{marginal probability graph} $G$. For a pair $(i,j)$ of alterations, let $p(i,j)$ denote the posterior probability that $i$ and $j$ are found in the same set.  We compute $p(i,j)$ using the samples from the MCMC algorithm (See Supplementary Section \ref{sec:mcmc-alg}).

Let $G = (V,E)$ be a complete, undirected weighted graph whose vertices are the alterations and where each edge $e \in E$ connects a pair of vertices $u, v$ with weight $p(u,v)$. Connected subgraphs of $G$ with many high-weight edges are the most exclusive alteration sets in $A$. We identify these most exclusive alteration sets by first removing all edges with weight below a threshold $\delta$ (See Supplementary Section \ref{sec:parameter-selection}). Let $C(\delta)$ be the connected components of size $\ge 2$ in the resulting graph. The output of \OurAlgo{} is the $C(\delta)$ alteration sets. We choose connected components as the output -- as opposed to some other partition of the graph such as cliques -- in order to be able to identify other topologies such as overlapping pathways (alteration sets), where two sets of alterations are connected by a cut node.

\subsection{Statistical significance}
While the score $\Phi(\mathbf{M})$ measures our surprise of observing exclusivity within each of the sets in $\mathbf{M}$ conditional on the observed frequencies of each alteration, there are a large number of possible collections,  and thus we might observe a high score by chance.  We evaluate the statistical significance of the collection $\mathbf{M}$ by comparing to a null distribution of scores obtained on permuted alteration matrices $A$ with the sample and alteration frequencies (sums of rows and columns of $A$) fixed \cite{MEMo,Gobbi2014}. Let $\Phi^*$ be the minimum score obtained over $N$ permutations.  We use the collections $\mathbf{M}$ satisfying $\Phi(\mathbf{M}) \le \Phi^*$ (thus each such collection has $P$-value $ < \frac{1}{N}$) to compute the marginal probability graph.

\subsection{Simultaneous analysis of alterations and cancer subtypes}
\label{sec:construct_subtype_matrix}
An important confounding factor in identifying cancer pathways \emph{de novo} by analyzing exclusive alterations is that certain alterations primarily occur in particular cancer subtypes \cite{Verhaak2010}. If we analyze a mixed set of samples with multiple subtypes, these subtype-specific alterations will be mutually exclusive in the data, even if they are not in the same biological pathway. When the subtypes are known in advance, one solution is to analyze subtypes separately; unfortunately this reduces sample numbers, thus reducing power to identify combinations of alterations that are shared across subtypes.  \OurAlgo{} addresses this problem by adding one new ``subtype row'' to the alteration matrix $A$ for each subtype. This subtype row contains an alteration in all samples \emph{excluding} those of the given subtype. Thus, the sets of alterations that are surprisingly exclusive with these subtype rows are the ones primarily \emph{altered} in that subtype. Note that when running \OurAlgo{} with subtype rows, we do not allow multiple subtypes to be placed in the same set. Because \OurAlgo{} simultaneously analyzes multiple alterations sets, \OurAlgo{} can identify exclusive sets containing subtype-specific alterations, general alterations, or any combination of these.

When analyzing the cancer dataset that included sample subtype classifications, we perform two runs of \OurAlgo{}. First we ran \OurAlgo{} on the alteration matrix $A$. Then we ran \OurAlgo{} on the alteration matrix with ``subtype rows'' as we described. We summarize the ensemble of statistically significant collections sampled by the MCMC algorithm in the two \OurAlgo{} runs by normalizing and combining the sampling frequencies of each collection across the two runs, and then computing the marginal probability graph from on the merged collection.

\subsection{Comparison to MEMo}
\label{sec:comparison-to-memo}
The MEMo algorithm \cite{MEMo} uses a permutation test to approximate the probability of observing exclusive mutations in a gene set $M$ with contingency table $X$. The permutation test works by permuting the rows in $A$ corresponding to the genes in $M$, and then determining if the permutation has a higher test statistic than $M$. This is then repeated $N$ times to obtain an empirical $P$-value. 

The crucial difference between MEMo and \OurAlgo{} is that 
%uses the exclusivity $T(X)$ as the test statistic, while  
MEMo uses the coverage $\Gamma(M)$ as the test statistic, while \OurAlgo{} uses the test statistic $T(X)$.  (For ease of exposition, let $\Gamma(X)$ also be defined a the coverage for a contingency table $X$.) The reasoning behind using the coverage as the test statistic is the idea that a gene set with mutually exclusive mutations will also have the highest coverage possible, for fixed frequencies of individual mutations.  While this is true for pairs of genes (which follows from the fact that $2 \times 2$ contingency tables have only one degree of freedom), when one examines three or more genes, maximizing coverage is not the same as maximizing exclusivity.  In fact, we can see that for a given contingency table $X$ it is possible to find another contingency table $X'$ with the same margins (gene frequencies) as $X$, but that has:
\begin{enumerate}
\item Higher exclusivity ($T(X') > T(X)$) and lower coverage ($\Gamma(X') < \Gamma(X)$), which could result in a deflated $P$-value for MEMo.
\item Lower exclusivity ($T(X') < T(X)$) but the same coverage ($\Gamma(X') = \Gamma(X)$), which would result in an inflated $P$-value for MEMo.\footnote{We have not found a case where $T(X') < T(X)$ and $\Gamma(X') > \Gamma(X)$, and conjecture that such a case does not exist.}
\end{enumerate}
See examples of both cases in Figure~\ref{fig:memo-comparison}.

\section{Results}
\label{sec:results}

\subsection{Visualization of results}
\label{sec:viz}
We created a web application for interactive visualization of the \OurAlgo{} results (\url{http://compbio-research.cs.brown.edu/comet}; See Figure~\ref{fig:website}). For each dataset, the website shows the modules in the \OurAlgo{} marginal probability graph. Users can change the minimum edge weight parameter $\delta$, which dynamically updates the modules. Edges in each module are labeled with the marginal probability. Users can view the rows of the alteration matrix that correspond to a given module, and also view, sort, and search through the collections sampled by \OurAlgo{} that include alterations in a given module.

\subsection{Benchmarking and comparison to other methods}
We compared \OurAlgo{} on two simulated mutation datasets to three other published methods  for finding mutually exclusive gene sets: Dendrix~\cite{Dendrix}, Multi-Dendrix~\cite{MultiDendrix} and muex~ \cite{muex}. In addition, we performed a separate comparison to MEMo \cite{MEMo} (See details in Section \ref{sec:comparison-to-memo}).

\begin{figure*}[h!]%{1\textwidth}
\begin{center}
  \includegraphics[width=\textwidth]{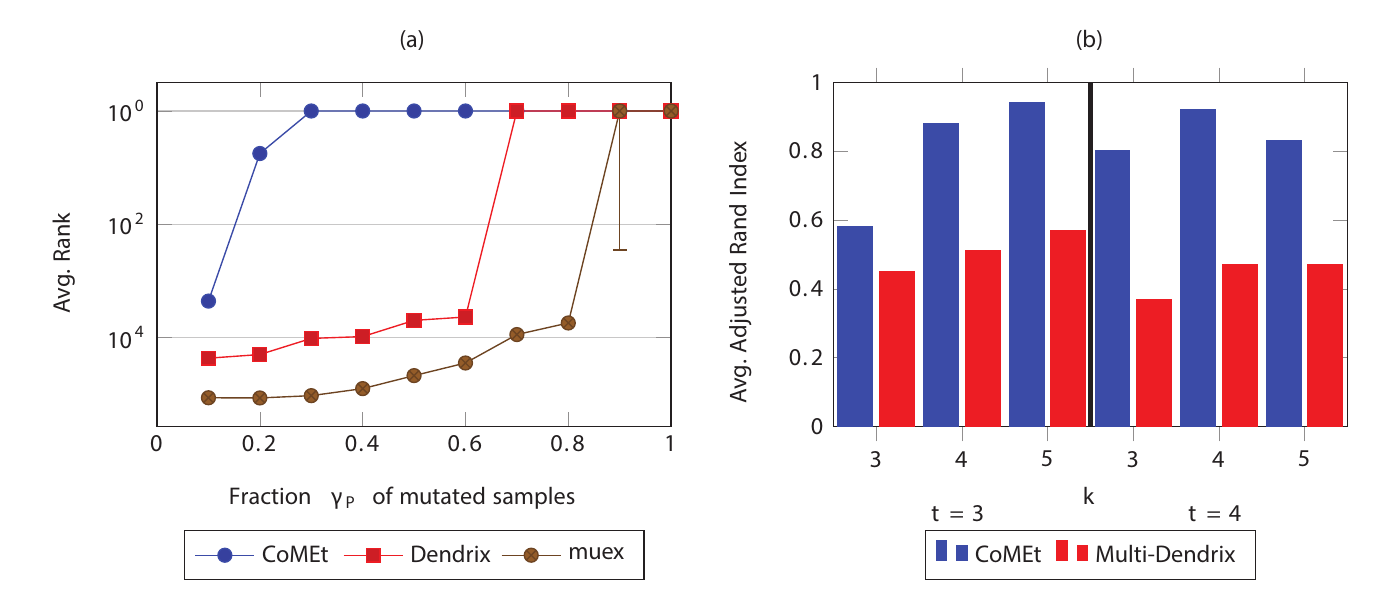}
	
    \caption{\textbf{Comparison of \OurAlgo{} with other methods on simulated data with $n=500$ samples.} (a) The average rank of the implanted pathway in the results from by \OurAlgo{} (blue), Dendrix (red), and muex (brown) in 25 simulated datasets as a function of coverage. The number of ties (i.e. gene sets with the same weight as the implanted pathway) are shown as error bars. (b) Comparison of \OurAlgo{} and Multi-Dendrix in identifying an implanted \emph{collection} containing multiple sets of alterations.  Bars indicate average of adjusted Rand index between reported and implanted collection across 25 simulated datasets.}
    \label{fig:sims}
    \end{center}
\end{figure*}

\subsubsection{Benchmarking of exclusivity scores for individual gene sets}
\label{sec:weight-comparison}
We first compared the exclusivity scores used by \OurAlgo{}, Dendrix, and muex for individual gene sets of size $k$ on simulated datasets that represent key features of cancer sequencing data.  In particular, each simulated dataset contains: (1) one implanted pathway $P$ with $k = 3$ genes that is altered in a fraction $\gamma_P$ samples with highly exclusive mutations; (2)  a set $C$ of 5 highly altered genes whose alterations are not necessarily exclusive; (3) other genes containing only passenger mutations that were altered at rate $q$.  The set $C$ models the highly recurrently altered genes that often appear in real cancer data sets, and can confound methods for identifying exclusive mutations.  Further details of the simulation are in the Appendix (See details in Supplementary Section \ref{sec:simulated-data}). 

We computed the average rank of the implanted pathway across 25 simulated datasets with $n=500$ samples, varying the coverage $\gamma_P=0.1, 0.2, \dots, 1.0$. We ran muex with the same parameters used in \cite{muex}, and ranked gene sets ascending by $P$-value. We then compared the algorithms by the average rank of the implanted pathway. We chose to use the average rank rather than a alternate measure such as the true positive or false positive rate because our simulated datasets only include a single true positive (the implanted pathway). We note that while we were able to reproduce the muex results from \cite{muex} using these parameters (See details in Section \ref{sec:comparison-real-data}), it is possible that different parameters would improve the performance of muex on this dataset.

We find that on average, \OurAlgo{} ranked the implanted pathway higher than the other methods for each coverage (Figure~\ref{fig:sims}).  For coverage $\gamma_P \ge 0.3$, \OurAlgo{} always ranked the implanted pathway first, while Dendrix ranked first only for $\gamma_P \ge 0.7$ and muex for $\gamma_P \ge 0.9$. Even with extremely low coverage of $\gamma_P = 0.1$ and 0.2, \OurAlgo{} was over an order of magnitude better than the other approaches. We also performed this comparison using both a smaller and larger number of samples ($n=250$ and $n=750$, respectively). We found that \OurAlgo{} improved as $n$ increased, as larger $n$ gave the \OurAlgo{} probabilistic test increased power, while Dendrix did not improve and muex still performed poorly (See Figure~\ref{fig:varying-size-sims}). We also compared the average runtimes of each weight function across all gene sets in each simulated dataset (Figure~\ref{fig:sim-runtimes}), finding that \OurAlgo{} ($<4$ minutes) and Dendrix ($<1$ minute) were much faster than muex (often an hour or more). This comparison demonstrates the superiority of the statistical score used in \OurAlgo{}, which is able to identify a pathway with low coverage alterations, even in datasets with highly recurrently mutated genes and many passenger mutations, and also runs in reasonable time.

\subsubsection {Benchmarking identification of collections of gene sets}
\OurAlgo{} and Multi-Dendrix~\cite{MultiDendrix} are the only available methods that simultaneously find collections containing more than one mutually exclusive set.  Thus, we compared these algorithms on simulated  datasets with overlapping and non-overlapping implanted gene sets   We generated simulated data using a procedure similar to above with three important differences. First, we implanted a collection $\mathbf{P}=(P_1, P_2, \dots, P_t)$ of $t$ pathways, each with exclusive mutations with total coverage $\gamma_\mathbf{P}$. Second, all genes in each implanted pathway are mutated in the same number of samples. Third, we include $m=20,000$ genes and remove those mutated in fewer than $1\%$ of total samples (Figure~\ref{fig:simulation_cutoff}). We generated datasets varying $t$ from 2 to 4 and $k$ from 3 to 5 with coverages $\gamma_P$ between 0.40 and 0.70 (See Table~\ref{tbl:sim-coverages}). We also generated datasets with \emph{overlapping} implanted pathways with $t=3$, $k$ from 3 to 5, with $\gamma_\mathbf{P}=(0.75, 0.75, 0.60)$.

On each  dataset, we ran \OurAlgo{} using $k=4, t=3$, and Multi-Dendrix using its default parameters of $t$ ranging from 2 to 4, and $k$ ranging from 3 to 5. We compared the consensus sets output by Multi-Dendrix with the modules output by \OurAlgo{}, using the adjusted Rand index (ARI) \cite{hubert1985comparing}, to score how well each algorithm identified the implanted pathways. The ARI measures the agreement between two partitions, with ARI = 1 indicating that two partitions are identical and ARI = -1 indicating that two partitions are maximally dissimilar. 
%More details about the adjusted Rand index can be found in \cite{hubert1985comparing}. 
\OurAlgo{} outperformed Multi-Dendrix in 11/12 simulated datasets (each containing 25 replicates) (Figure~\ref{fig:sims}b and Table~\ref{tbl:collection-sims}).  \OurAlgo{} found a much larger fraction of the implanted pathways (difference in ARI was $>0.2$ for 8/12 datasets). Furthermore, \OurAlgo{} had an ARI $>0.5$ for all 12 datasets, and ARI$>0.8$ for 7/12 datasets.  We emphasize that we ran \OurAlgo{} with a \emph{single} value of $t$ and a \emph{single} value of $k$ over all datasets even though the size and number of implanted pathways varied across datasets.  In contrast, Multi-Dendrix was run with a range of parameter values.  This demonstrates that \OurAlgo{} is much less sensitive to parameter choices than Multi-Dendrix.

We also compared the output of \OurAlgo{} and Multi-Dendrix using the true values of $t$ and $k$.  We found that \OurAlgo{} outperformed Multi-Dendrix on 11/12 datasets (Table~\ref{tbl:collection-sims}). This shows that the statistical score used by \OurAlgo{} and the MCMC sampling are important features, even on simulated datasets where the implanted collections are fairly strong signals in the data.

\begin{table*}[h!]
\footnotesize
\centering
\begin{tabular}{ @{}llrllr@{} }
\toprule
\multicolumn{3}{c}{muex} & \multicolumn{3}{c}{\OurAlgo{}} \\ 
\cmidrule(l){1-3}\cmidrule(l){4-6}
Alterations set & $\Phi(M)$ & Statistic & Alteration set & $\Phi(M)$ & Statistic \\ 

\cmidrule(l){1-6} 
\multicolumn{6}{c}{Multi-Dendrix GBM dataset \cite{MultiDendrix}, $k=3$, $t=3$}\\
\cmidrule(l){1-6} 
 \textbf{\textit{EGFR}},\textbf{\textit{PDGFRA}}(A),\textbf{\textit{PTEN}}(D) & 0.0029 & 1.89 &
 \textbf{\textit{CDK4}}(A),\textbf{\textit{CDKN2A}}(D),\textbf{\textit{RB1}} & $6.3\times10^{-19}$ & -3.93\\
 \textit{FRMPD4}(D),\textbf{\textit{MDM2}}(A),\textbf{\textit{PIK3CA}} & 0.011 & 2.63 &\textbf{\textit{CDKN2A}}(D),\textbf{\textit{MDM2}}(A),\textbf{\textit{TP53}} & $4.8\times10^{-17}$ & -5.38\\
 \textit{ABP1},\textbf{\textit{ARID2}}(D),\textit{DUSP27} & 0.18232  & 1.78 & 
 \textbf{\textit{IDH1}},\textbf{\textit{PTEN}},\textbf{\textit{PTEN}}(D)& $1.1\times10^{-8}$& -0.55\\
 \cmidrule(l){1-6} 
 \multicolumn{6}{c}{muex GBM dataset \cite{muex}, $k=4$, $t=3$} \\
 \cmidrule(l){1-6} 
 \textit{ABCC9},\textbf{\textit{PIK3CA}},\textbf{\textit{RPL5}},\textit{TRAT1} & 0.96 & 2.62 &\textbf{\textit{EGFR}}, \textit{GCSAML}, \textbf{\textit{IDH1}}, \textit{OTC} & $1.9\times10^{-8}$ & -14.72\\ 
 \textit{CNTNAP2},\textbf{\textit{IDH1}},\textit{KEL},\textit{SCN9A} & 0.11 & 6.35 &
  \textit{ABCC9},\textbf{\textit{CDK4}},\textbf{\textit{CDKN2B}},\textbf{\textit{RPL5}} & $1.1\times10^{-7}$  & -8.01\\
 \textit{CDH18},\textit{MMP13},\textit{SULT1B1},\textit{TRIM51} & 0.32  & 3.24 &
  \textbf{\textit{CDK4}},\textit{CNTNAP2},\textbf{\textit{NF1}},\textit{SCN9A} & $5.9\times10^{-5}$ & -2.51\\
 \bottomrule
\end{tabular}
\caption{\textbf{Collections of alterations reported by muex and \OurAlgo{} on the Multi-Dendrix GBM dataset \cite{MultiDendrix} (with $k=3$ and $t=3$), and on the muex GBM dataset \cite{muex} (with $k=4$ and $t=3$)}.  Bolded alterations indicate genes in the COSMIC Cancer Census \cite{COSMIC}.
}
\label{tbl:muex_comp}
\end{table*}

\subsubsection {Comparison to muex on real data}
\label{sec:comparison-real-data}
We compared \OurAlgo{} to muex \cite{muex} using two different versions of the TCGA glioblastoma (GBM) dataset: (1) the dataset from \cite{MultiDendrix} containing $398$ alterations and $261$ samples; (2) the dataset from \cite{muex}, containing $83$ alterations and $236$ samples (See Section \ref{sec:data-gbm}). There are 184 samples in both the Multi-Dendrix GBM and muex GBM datasets. Besides the samples, the main difference between these two datasets is that the muex dataset is restricted to only 83 significantly recurrent alterations.

Since the muex score is for single alteration sets, we ran muex iteratively to identify collections of alteration sets.  That is, we run muex to find the top scoring alteration set, remove those alterations, and repeat $t-1$ times. We ran muex with the parameters used in \cite{muex}, restricting to alteration sets with coverage at least $0.3$, impurity lower than $0.5$, and a significance cutoff of $0.05$.  On the muex GBM dataset, we ran \OurAlgo{} and muex with $k=4$ and $t=3$ to match the parameters used in \cite{muex}. On the Multi-Dendrix GBM dataset, we ran \OurAlgo{} and muex  with $k=3$ and $t=3$, since muex aborted with an out-of-memory error for  $k=4$ on this dataset.

On both GBM datasets, \OurAlgo{} identifies collections with much more significant exclusivity.  Moreover, more of the genes in the \OurAlgo{} collections are known cancer genes (according to the COSMIC Cancer Census \cite{COSMIC}) compared to the genes in the muex collections (Table~\ref{tbl:muex_comp}). On the Multi-Dendrix GBM dataset, \OurAlgo{} identifies three collections that overlap the Rb (\textit{CDK4}, \textit{CDKN2A}, \textit{RB1}), p53 (\textit{TP53}, \textit{MDM2}, \textit{CDKN2A}), and PI(3)K (\textit{PTEN}, \textit{IDH1}) signaling pathways. Each of these sets include surprisingly exclusive alterations, with $\Phi(M)$ ranging from $10^{-8}$ to $10^{-19}$, and all the alterations are in cancer genes. In contrast, muex identifies sets with lower coverage and less surprising exclusivity, with $\Phi(M) > 10^{-3}$ for each set, and three of the alterations are not in known cancer genes. 

On the muex GBM dataset, \OurAlgo{} again identifies more exclusive alteration sets that overlap more known cancer genes, while muex reports few known cancer genes with most having an uncertain association with cancer.  In general this dataset seems to include more spurious alterations, as both algorithms identify less exclusive sets with fewer cancer genes than on the Multi-Dendrix GBM dataset. This might be a result of the different handling of copy number aberrations in the two papers (see \cite{MultiDendrix} and \cite{muex}).

\begin{table*}[h!]
\footnotesize
\centering
\begin{tabular}{ @{}llrllr@{} }
\toprule
\multicolumn{3}{c}{Multi-Dendrix} & \multicolumn{3}{c}{\OurAlgo{}} \\ 
\cmidrule(l){1-3}\cmidrule(l){4-6}
Alterations set & $\Phi(M)$ & W(M) & Alteration set & $\Phi(M)$ & W(M) \\ 
\cmidrule(l){1-6} 
\multicolumn{6}{c}{Pan-cancer GBM dataset \cite{leiserson2014pan}, $k=3$, $t=4$}\\
\cmidrule(l){1-6} 
\textit{CDKN2A}(D),\textit{CDK4}(A),\textit{RB1} & $1.4\times10^{-13}$ & 160 &
\textit{CDKN2A}(D),\textit{CDK4}(A),\textit{RB1} & $1.4\times10^{-13}$ & 160  \\
\textit{TP53},\textit{MDM2}(A), \textbf{\textit{MDM4}}(A) & $3.7\times10^{-6}$ & 128 &
\textit{TP53},\textit{MDM2}(A),\textit{STAG2} & $3.8\times10^{-7}$ & 119\\
\textit{PTEN},\textbf{\textit{PIK3CA}},\textit{IDH1} & $3.5\times10^{-4}$  & 125 & 
\textit{PTEN},\textit{LRP1B},\textit{IDH1} & $6.9\times10^{-5}$  & 112 \\
\textit{EGFR},\textit{NF1},\textit{PDGFRA}(A) & $1.1\times10^{-2}$  & 120 & 
\textit{EGFR},\textit{NF1},\textbf{\textit{CALCR}} & $4.3\times10^{-4}$  & 111 \\
 \cmidrule(l){1-6} 
\multicolumn{6}{c}{Pan-cancer GBM dataset \cite{leiserson2014pan} without MutSigCV filter, $k=3$, $t=4$}\\
\cmidrule(l){1-6} 
\textit{CDKN2A}(D),\textit{CDK4}(A),\textit{RB1} & $1.4\times10^{-13}$ & 160 &
\textit{CDKN2A}(D),\textit{CDK4}(A),\textit{RB1} & $1.4\times10^{-13}$ & 160  \\
\textit{TP53},\textit{MDM2}(A),\textbf{\textit{EGFR}} & $3.6\times10^{-4}$ & 144 &
\textit{TP53},\textit{MDM2}(A),\textit{STAG2} & $3.8\times10^{-7}$ & 119 \\
\textit{PTEN},\textbf{\textit{MUC16}},\textit{IDH1} & $1.3\times10^{-3}$  & 130 & 
\textit{PTEN},\textit{LRP1B},\textit{IDH1} & $6.9\times10^{-5}$  & 112 \\
\textbf{\textit{TTN}}, \textbf{\textit{PIK3R1}},\textit{PDGFRA}(A) & $2.5\times10^{-1}$   & 109 & 
\textit{EGFR},\textit{NF1},\textbf{\textit{PKHD1}} & $1.1\times10^{-4}$  & 117 \\
 \bottomrule
\end{tabular}
\caption{\textbf{Collections of alterations reported by Multi-Dendrix and \OurAlgo{} on Pan-Cancer glioblastoma data \cite{leiserson2014pan} (with $k=3$ and $t=4$). }Bolded alterations indicate differences between alteration datasets with and without MutSigCV filter.} 
\label{tbl:mdendrix_comp_gbm}
\end{table*}
 
\subsubsection {Comparison to Multi-Dendrix on real data }
Because \OurAlgo{} conditions on the observed alteration frequencies, we argue that it is less biased towards gene that have high mutation frequencies because of their higher background mutation rates; e.g. long genes.  
To illustrate this point, we compare \OurAlgo{} with Multi-Dendrix  on Glioblastoma (GBM) and Breast cancer  (BRCA) with and without the MutSigCV \cite{Lawrence2013} filter that requires that frequently mutated genes have low MutSigCV $q$-values (See Section \ref{appendix:mutation_data} for details).
We ran \OurAlgo{} and Multi-Dendrix with $k=3$ and $t=4$ on GBM and $k=4$ and $t=4$ on BRCA. We used mutation data from the TCGA Pan-Cancer dataset \cite{leiserson2014pan} which contains whole-exome and copy number array data, and downloaded MutSigCV output from the corresponding Synapse repository (syn2812925). We used different TCGA GBM and BRCA datasets here than we present in Section \ref{sec:results} because of the availability of MutSigCV results on the Pan-Cancer dataset.
For each cancer, we generated two datasets. In one dataset, we applied a MutSigCV filter to remove highly altered genes (altered in $>2.5\%$ of samples) but insignificant by MutSigCV ($q$-value $< 0.1$). The second dataset did not include any MutSigCV filter. 

We found that \OurAlgo{} produces almost identical results when using the alteration dataset with or without the MutSigCV filter, in both GBM and BRCA (See Table \ref{tbl:mdendrix_comp_gbm} and Supplementary Table \ref{tbl:mdendrix_comp_brca}).  In contrast, the  Multi-Dendrix results are different with and without the MutSigSV filter. Without the MutSigCV filter, Multi-Dendrix output includes highly altered genes in GBM (including \textit{TTN} and \textit{MUC16}) that are known to have high background mutation rates. Furthermore, Multi-Dendrix gives the collection including \textit{TTN} and \textit{MUC16} higher weight $W'$ than the highest weight collection obtained with the MutSigCV filter. Multi-Dendrix also has quite different results between the datasets with and without the MutSigCV filter in BRCA, while \OurAlgo{} is largely consistent. This observation demonstrates that genes with high alteration frequencies can dominate the mutual exclusivity signal in Dendrix weight function $W(M)$ and bias the algorithms towards identifying gene sets where the majority of the coverage comes from one gene, while \OurAlgo{} gives less weight to these genes, which are likely not cancer genes.

\subsection {\OurAlgo{} results on real cancer datasets}
\label{sec:real-data}
We ran \OurAlgo{} on four mutation datasets from TCGA: glioblastoma (GBM) \cite{TCGAGBM2008}, breast cancer (BRCA) \cite{TCGABRCA}, gastric cancer (STAD) \cite{TCGASTAD} and acute myeloid leukemia (AML) \cite{TCGAAML}. Because \OurAlgo{} can analyze any type of binary alterations, we include many types of alterations in these datasets: small indels and single nucleotide variations, copy number aberrations, aberrant splicing events, gene fusions, and (for BRCA and STAD) cancer subtype.  See Section \ref{appendix:mutation_data} for details on these datasets and Supplementary Section \ref{sec:parameter-selection} for details on parameters.

\begin{figure*}[h!]%{1\textwidth}
\begin{center}
  \includegraphics[width=\textwidth]{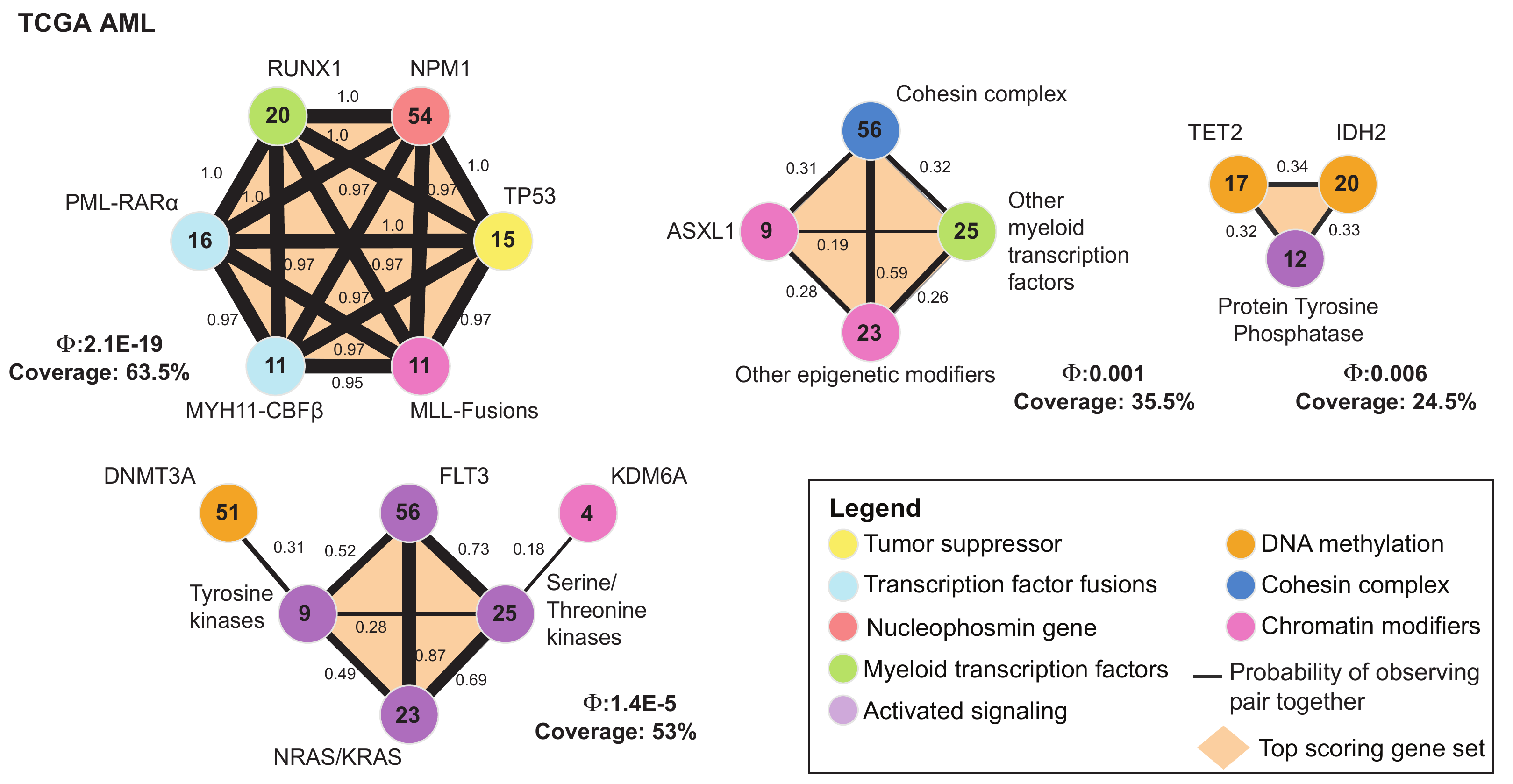}
  \caption{\textbf{\OurAlgo{} results on TCGA AML.} 
  Each circle represents the alterations in a gene or genomic region with the number in the circle indicating the number of samples in which the alteration occurs. Black lines are edges in the marginal probability graph with indicated probabilities. Orange polygons indicate the sets in the collection $\mathbf{M}$ with the most significant value $\Phi(\mathbf{M})$. $\Phi$ values are shown for each top set (orange polygon). 
  }
  \label{fig:real-data-results-aml}
\end{center}
\end{figure*}

\paragraph{Acute myeloid leukemia (AML)}
\label{sec:aml}
We first ran \OurAlgo{} with $t = 4$ alteration sets, each of size $k=4$.  The \OurAlgo{} output contains four mutually exclusive modules that include $18$ alterations (Figure~\ref{fig:aml-results}). These $4$ modules are: (1) \textit{TP53}, \textit{RUNX1}, \textit{NPM1}, \textit{PML-RAR$\alpha$} ($52.5\%$ of samples); (2) \textit{KDM6A}, \textit{FLT3}, tyrosine kinases, \textit{RAS} proteins, serine/threonine kinases, \textit{DNMT3A}, \textit{MLL}-X fusions, \textit{MYH11-CBF$\beta$}, and \textit{RUNX1-RUNX1T1} fusion ($70\%$ of samples); (3) cohesin complex, other myeloid transcription factors, and other epigenetic modifiers ($33\%$ of samples); (4) \textit{TET2} and \textit{IDH2} ($18.5\%$ of samples). 

The recent TCGA AML publication~\cite{TCGAAML} reported strong mutual exclusivity (using an earlier version of \OurAlgo{} algorithm, called Dendrix++) across several expert-defined classes.
Thus, we increased the value of $k$ compute $t = 4$ gene sets with sizes $k = 6, 4, 4, 3$.  Because of the larger values of $k$, we increased the number of MCMC iterations to $200$ million. The resulting marginal probability graph ($\delta =0.179$) contained $4$ mutually exclusive modules with a total of $19$ genes (Figure~\ref{fig:real-data-results-aml}). 

The first module contains six perfectly mutually exclusive alterations.  These six alterations include: mutations in \textit{TP53}, \textit{RUNX1}, \textit{NPM1}; \textit{PML-RAR$\alpha$}, \textit{MYH11-CBF$\beta$} fusion genes, and other \textit{MLL} fusions, which we denote as  \textit{MLL}-X fusions, following \cite{TCGAAML}.
These six alterations are known to be drivers in AML, and together are found in $63.5\%$ of the samples. 
%\hsinta{ \textit{TP53}, \textit{RUNX1} and \textit{NPM1} are transcriptional regulators. \textit{PML-RAR$\alpha$} and \textit{MYH11-CBF$\beta$} are transcription-factor fusion genes}; 
These fusion genes are defining aberrations for certain subtypes of AML, as \textit{PML-RAR$\alpha$}, \textit{MYH11-CBF$\beta$}, and \textit{MLL}-fusions are associated with acute promyelocytic leukemia, acute monoblastic or monocytic leukemia, and acute megakaryoblastic leukemia, respectively.
The second module (altered in $63\%$ of samples) contains receptor tyrosine kinases (RTKs) and their downstream RAS target proteins.  These include mutations in the \textit{FLT3} tyrosine kinase, other tyrosine kinases, serine/threonine kinases, and \textit{RAS} proteins.  Two additional genes \textit{DNMT3A} and \textit{KDM6A}, are also included in this set.  These genes are involved in DNA/histone methylation, and their interactions with the other RTK/RAS genes in the set are less clear.  Notably, the marginal probability graph (Figure~\ref{fig:real-data-results-aml}) shows that the connection between \textit{DNMT3A} and other genes in the set is largely due to its mutual exclusivity with other tyrosine kinases, and in fact a number of samples have mutations in both \textit{FLT3} and \textit{DMNT3A} (Figure~\ref{fig:mutation-matrices}).  Thus, the patterns of exclusivity/co-occurrence between alterations may be subtle, demonstrating the advantages of \OurAlgo{}'s approach to simultaneously examine multiple collections of sets of alterations.

The third module (altered in $35.5\%$ of samples) contains genes related to chromatin modification and gene regulation including  \textit{ASXL1}, the cohesin complex, other myeloid transcription factors, and other epigenetic modifiers.  Finally, the fourth module (altered in $24.5\%$ of samples) contains genes related to DNA methylation including \textit{TET2}, \textit{IDH2} and protein tyrosine phosphatases. 
Mutual exclusivity between \textit{TET2} and \textit{IDH2} in AML has been previously reported \cite{figueroa2010leukemic, abdel2013mutations, metzeler2011tet2}. Moreover, recent work provides a mechanistic explanation for this observed exclusivity: Figeroa et al. \cite{figueroa2010leukemic} show that mutant \textit{IDH1\/2} inhibits \textit{TET2}'s function in demethylation of 5-methylcytosine.
%hydroxylation reaction of methylcytosine.}
These results demonstrate that \OurAlgo{} is able to extract multiple functional modules directly from alteration data.

\begin{figure*}[h!]%{1\textwidth}
\begin{center}
  \includegraphics[width=\textwidth]{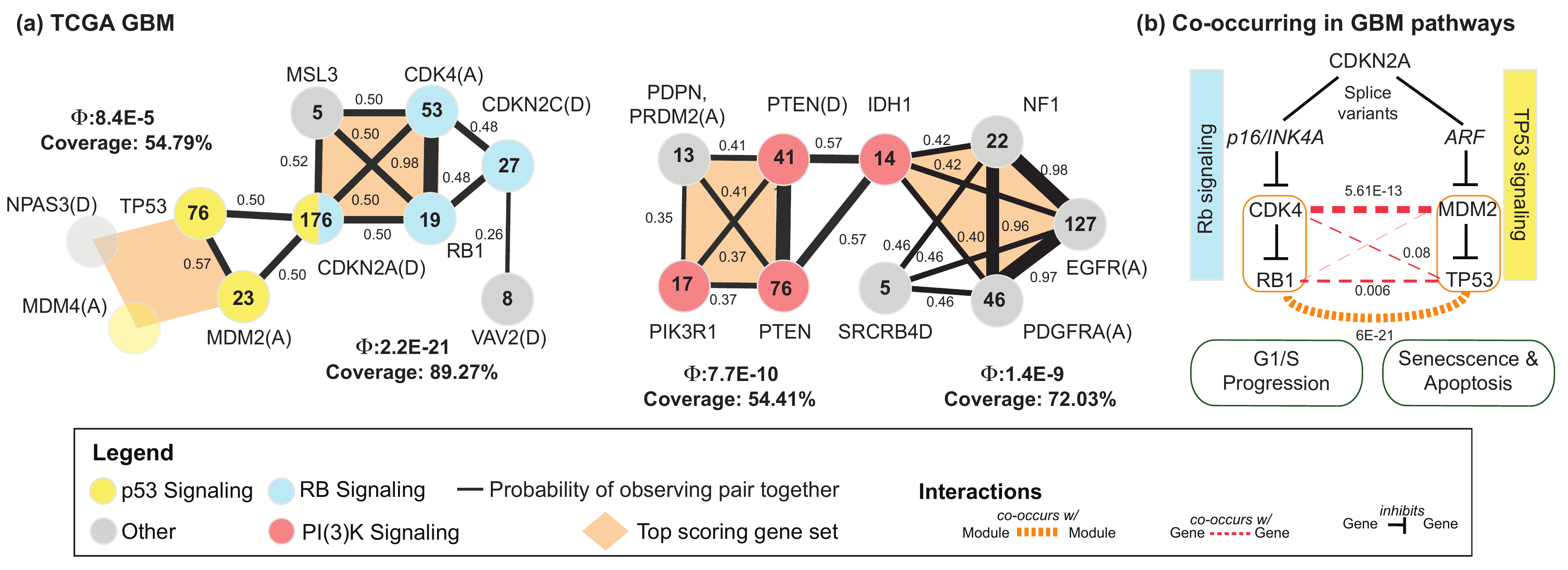}
  \caption{\textbf{\OurAlgo{} results on (a) TCGA GBM.} 
  Style is the same as in Figure~\ref{fig:real-data-results-aml}, except for the addition of character(s) inside parentheses after a gene name indicates the type of alterations in the gene: $\mathbf{D}$: deletion, $\mathbf{A}$: amplification.
%  \ben{[The coloring of nodes and correspondence with legend is not described.]}
(b) Different splice variants of \textit{CDKN2A} are part of both the Rb signaling (left) and p53 signaling (right) pathways. \OurAlgo{} recovers this relationship as two separate exclusive gene sets. The Rb signaling (\textit{RB1} and \textit{CDK4}) and p53 signaling (\textit{MDM2} and \textit{TP53}) gene sets (not including \textit{CDKN2A}) exhibit a statistically significant number of co-occurring mutations ($P=6\times10^{-21}$), where the co-occurrence between the pairs (dotted orange line) is  more significant than between any of the pairs of genes (dotted red line).
  }
  \label{fig:real-data-results-gbm}
\end{center}
\end{figure*}

\paragraph{Glioblastoma multiforme (GBM)}
\label{sec:gbm}
We ran \OurAlgo{} on the TCGA GBM dataset from \cite{MultiDendrix} with $t=4$ and $k=4$. While \cite{MultiDendrix} removed amplifications in \textit{EGFR} because they were so frequent it confounded their analysis, we added these amplifications back when running \OurAlgo{}, treating \textit{EGFR} amplifications and \textit{TP53} as subtypes so they could not be sampled in the same set (See Section \ref{sec:construct_subtype_matrix} for details). The resulting marginal probability graph ($\delta= 0.263$) includes two mutually exclusive modules (Figure~\ref{fig:real-data-results-gbm}(a)). 
 
The first module includes: three genes in the Rb signaling pathway (\textit{CDK4}, \textit{RB1}, \textit{CDKN2C}) and three genes in the p53 signaling pathway (\textit{TP53},  \textit{MDM2}, and \textit{MDM4})), as annotated by the original TCGA GBM publication \cite{TCGAGBM2008}.  This module also contains deletions in \textit{CDKN2A}, which is a member of both the Rb signaling and p53 signaling pathways.
Indeed, it is well known that different isoforms of the \textit{CDKN2A} gene are involved in the Rb and p53 signaling pathways (See Figure~\ref{fig:real-data-results-gbm}(b) and also \cite{TCGAGBM2008}) and the genomic deletion of \textit{CDKN2A} affects both isoforms.  
Moreover, we find that the pairs \textit{CDK4}-\textit{RB1} and \textit{MDM2}-\textit{TP53} have surprisingly co-occurring alterations ($P=6\times10^{-21}$; See Figure~\ref{fig:real-data-results-gbm}(b)).  This co-occurrence is stronger than the co-occurrence of alterations in individual genes. This pattern indicates that glioblastomas can alter the function of the Rb and p53 signaling pathways either by deleting \textit{CDKN2A}, \emph{or} by altering one gene in each of the pairs (\textit{CDK4}, \textit{RB1}) and (\textit{TP53}, \textit{MDM2}).  We emphasize that \OurAlgo{} identified this overlapping module by sampling \emph{non-overlapping} exclusive sets.  Finally, this module contains alterations in three additional genes: \textit{NPAS3}, \textit{VAV2}, and \textit{MSL3}.  \textit{NPAS3} has been studied as a novel late-stage acting progression factor in gliomas with tumor suppressive functions \cite{kamnasaran2010393, moreira2011npas3}.  \textit{VAV2} has been reported to regulate \textit{EGFR}, and knockdown of \textit{VAV2} enhanced EGFR degradation and further reduced cell proliferation \cite{thalappilly2010vav2}. \text{MSL3} is a member of the male-specific lethal (MSL) complex and is thought to play a role in transcriptional regulation. As reported in \cite{MultiDendrix}, the MSL complex also includes MOF, which regulates p53 in cell cycle and may be involved in cancer \cite{rea2007males}. We note that Multi-Dendrix identifies similar Rb and p53 signaling modules \cite{MultiDendrix}, with the important difference being that \OurAlgo{} correctly places  \textit{CDKN2A} in a module with both the Rb and p53 signaling pathways, consistent with the figure in the TCGA GBM publication \cite{TCGAGBM2008}.
 
The second module includes alterations in the PI(3)K signaling pathway -- including \textit{PIK3R1}, \textit{PTEN}, deletion of \textit{PTEN} and \textit{IDH1} -- as well as amplifications in the genes (\textit{EGFR}, \textit{PDGFRA}) and in a region containing \textit{PRDM2} and \textit{PDPN}.
Additional genes in this module are \textit{NF1} and \textit{SRCRB4D}.
The PI(3)K signaling pathway genes overlap the results reported by Multi-Dendrix on this dataset in \cite{MultiDendrix}, with the important differences being \OurAlgo{} includes \textit{NF1} and amplifications in \textit{EGFR} (which were not analyzed by \cite{MultiDendrix}). In this module, we also found one mutually exclusive gene set (from the highest weight collection) that includes \textit{EGFR}, \textit{IDH1}, \textit{NF1}, and \textit{PDGFRA}.  Alterations in these genes have strong association with individual subtypes in GBM \cite{Verhaak2010}:  \textit{EGFR} amplification is associated with the Classical GBM subtype, \textit{IDH1} and \textit{PDGFRA} amplification are associated with Proneural GBM subtype, and \textit{NF1} is associated with Mesenchymal GBM subtype. This shows that mutually exclusive gene sets can result from subtype-specific mutations.

Finally, \textit{SRCRB4D} is a scavenger receptor with no known associations with cancer.  However, two other scavenger receptor genes have previously reported roles in glioblastoma.  Homozygous deletions of \textit{DMBT1} were reported in glioblastomas and astrocytomas \cite{mollenhauer1997dmbt1,motomura2012dmbt1}.  \textit{CD36} was recently reported to be involved in cancer stem cell maintainence in glioblastoma \cite{hale2014cancer}.

These results show that \OurAlgo{} can automatically find large portions of the pathways that were manually curated in TCGA GBM publication \cite{TCGAGBM2008}, including overlapping pathways.  Moreover, \OurAlgo{} identifies additional genes with putative roles in glioblastoma and significant exclusivity with other known glioblastoma genes.

\begin{figure*}[h!]%{1\textwidth}
\begin{center}
  \includegraphics[width=\textwidth]{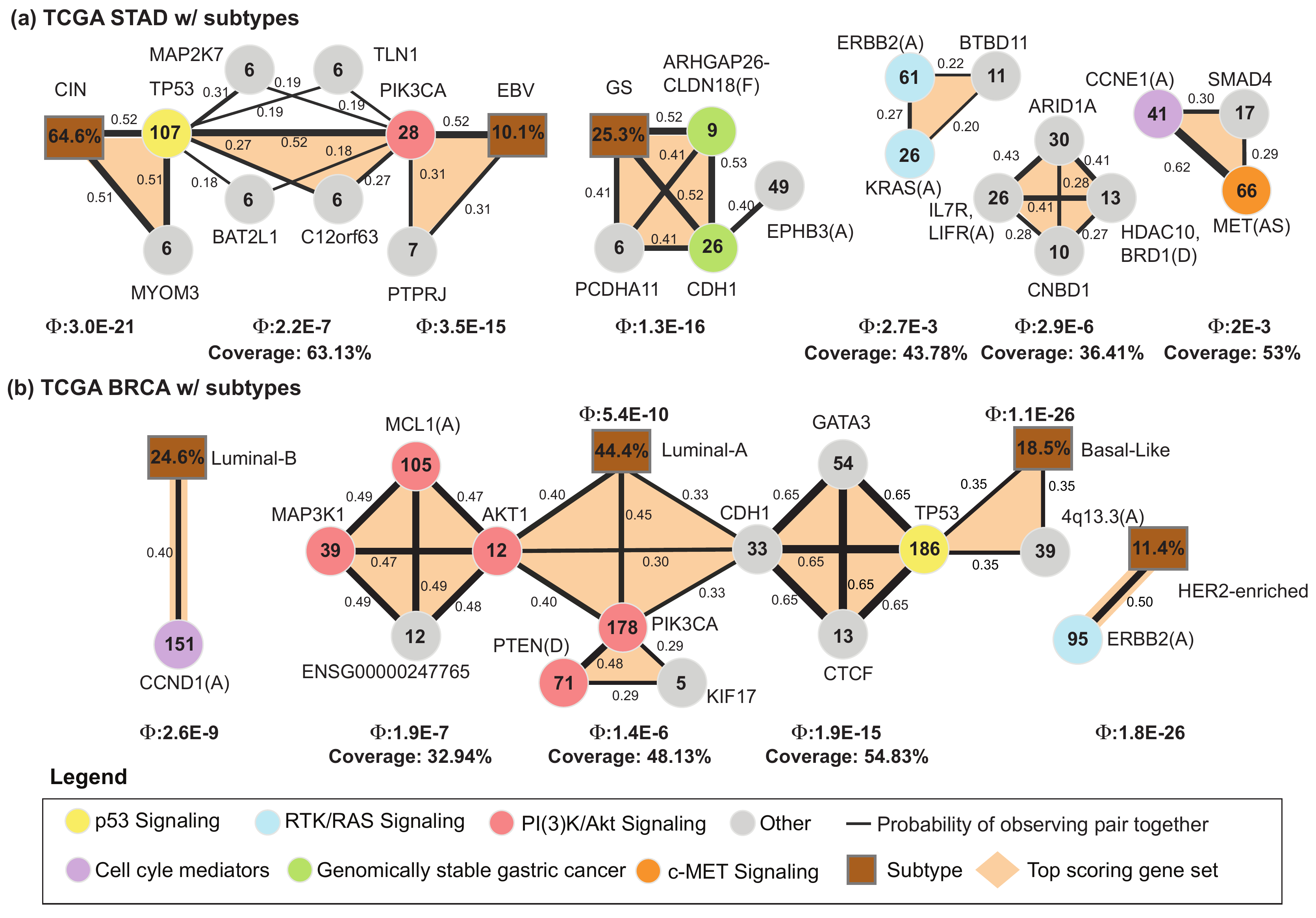}
  \caption{\textbf{\OurAlgo{} results on (a) TCGA STAD subtypes, (b) TCGA BRCA subtypes.}  Style is the same as in Figure~\ref{fig:real-data-results-gbm}, except for the addition of subtype alterations (brown) and that character(s) inside parentheses after a gene name indicates the type of alterations in the gene ($\mathbf{AS}$: alternative splicing event, $\mathbf{F}$: fusion gene). Here, an edge between a subtype and an alteration indicates that the alteration is associated with the subtype.
}
  \label{fig:real-data-results-subtypes}
\end{center}
\end{figure*}

\paragraph{Gastric cancer (STAD)}
We performed two runs of \OurAlgo{} on the TCGA gastric cancer (STAD) dataset, and then merged the runs (described in Section \ref{sec:construct_subtype_matrix}). We first ran \OurAlgo{} with $t=4$ and $k=4$. We then ran \OurAlgo{} on a STAD dataset that included sample subtype classifications. TCGA recently classified gastric cancers into four subtypes based on integration of different molecular data~\cite{TCGASTAD}.  To examine the relationships between subtypes and other alterations, we introduce ``subtype alterations" for the three subtypes from~\cite{TCGASTAD} (we did not include the hypermutated samples from the MSI subtype in our analysis). As described in Section \ref{sec:construct_subtype_matrix}, these ``subtype alterations" are marked as altered in samples that are \emph{not} members of the subtype, so that mutual exclusivity between an ``subtype alteration" and another alteration indicates that the alteration is enriched in the subtype. We ran \OurAlgo{} on the STAD dataset with subtype alterations using $k=4$ and $t=3$ (the number of subtypes).

\OurAlgo{} identified five mutually exclusive modules from the marginal probability graph  ($\delta= 0.132$) in the STAD dataset (Figure~\ref{fig:real-data-results-subtypes}(a)).  Each of these modules includes known cancer genes and novel candidate genes.  Two modules indicate subtype-specific altered genes and pathways.  The first module (altered in 69\% (150/217) of the STAD samples) includes two genes, \textit{TP53} and \textit{PIK3CA}, that are enriched for alterations in the CIN and EBV subtypes, respectively. TCGA gastric study reported that $80\%$ of EBV tumors contain an alteration in PIK3CA, and they suggested that EBV tumors might respond to PI3-kinase inhibitors \cite{TCGASTAD}. Given this strong signal, it is not surprising that these two genes appear in \OurAlgo{} results.  However, these signals do not dominate the  \OurAlgo{} results, and four other interesting modules are also output. There are six other mutated genes in this first module including \textit{MAP2K7}, \textit{TLN1}, \textit{BAT2L1}, \textit{C12orf63} (recently renamed \textit{CFAP54}), \textit{MYOM3} and \textit{PTPRJ}.  Given the rarity of these mutations, their significance is unclear.

The second STAD model includes the genomically stable (GS) subtype, mutations in \textit{CDH1}, mutations in \textit{PCDHA11}, \textit{ARHGAP6-CLDN18} fusions, and amplification of a region containing \textit{EPHB3}. \textit{CDH1} somatic mutations and \textit{ARHGAP6-CLDN18} fusions were reported to be mutually exclusive and enriched in the genomically stable subtype in gastric cancer~\cite{TCGASTAD}, and \OurAlgo{} recapitulates this result.  \textit{EPHB3} is the member of Eph/ephrin signaling which controls the compartmentalization of cells in epithelial tissues. A recent study \cite{solanas2011cleavage} demonstrated EphB receptors (e.g. \textit{EPHB1} and \textit{EPHB3}) interact with \textit{CDH1} in epithelial intestinal cells that regulates the formation of E-cadherin-based adhesions.  This interaction explains the perfect mutual exclusivity between \text{CDH1} and \textit{EPHB3}, which to our knowledge is the first report of this relationship. This demonstrates that mutual exclusivity between pairs of alterations/subtypes may have subtle explanations, further underscoring the need for analysis of collections of multiple alterations.

The third module (altered in $95/217$ of samples) includes amplifications of \textit{KRAS} and \textit{ERBB2}, and mutations in \textit{BTBD11}. \textit{KRAS} and \textit{ERBB2} are members of the RTK/RAS signaling pathway, and their role in cancer is well-documented. Little is known about the function of \textit{BTBD11}, and thus the significance of the mutations is unclear.

The fourth STAD module ($115/217$ of samples) contains three altered genes, including amplifications of \textit{CCNE1}, mutations in \textit{SMAD4} and splice-site mutations in \textit{MET}. \textit{CCNE1} is a well-known cell cycle mediator, and \textit{SMAD4} is a member of the TGF-$\beta$ pathway, and \textit{MET} participates in the RTK/RAS signaling pathway \cite{TCGASTAD}.

The fifth STAD module ($79/217$ of samples) contains four altered genes, including amplifications in a region with \textit{IL7R} and \textit{LIFR}, deletions in a region with \textit{HDAC10} and \textit{BRD1}, mutations in \textit{ARID1A}, and mutations in \textit{CNBD1}. \textit{ARID1A} is a well-known cancer gene shown to be significantly mutated in gastric cancer \cite{TCGASTAD}. Moreover, inhibition of \textit{HDAC10} has been reported with association with human gastric cancer cells~\cite{lee2010inhibition}.  Gain-of-function mutations in \textit{IL7R} have been reported to associated with childhood acute lymphoblastic leukemia \cite{shochat2011gain}.  Our \OurAlgo{} results suggest that \textit{IL7R} mutations may have a role in gastric cancer as well.

\paragraph{Breast cancer (BRCA)}
\label{sec:brca}
We performed two runs of \OurAlgo{} on the TCGA breast cancer (BRCA) dataset, and then merged the runs. We first ran \OurAlgo{} with $k=4$ and $t=4$. We then introduced subtype alterations for four subtypes from~\cite{TCGABRCA} (as described in Section \ref{sec:construct_subtype_matrix}). Breast cancers are traditionally classified into multiple subtypes based on mRNA expression. Here we analyze four subtypes: luminal A, luminal B, basal-like, and HER2-enriched.  We ran \OurAlgo{} on a BRCA dataset that included sample subtype classifications with $k=4$ and $t=4$ (the number of subtypes).

\OurAlgo{} identified three subtype-specific modules and three modules with mutated genes (Figure~\ref{fig:real-data-results-subtypes}(b)) in the marginal probability graph ($\delta= 0.287$).  The first module shows the strong association between amplification of \textit{CCND1} and the luminal B subtype as previously reported~\cite{holm2012ccnd1}. Similarly, the third module shows the strong association between \textit{ERBB2} amplification and the HER2 (\textit{ERRB2}) -enriched subtype.
 
 The second module shows a complicated relationship between: (1) subtype-associated alterations in the Luminal-A and Basal-Like subtypes, and (2) mutual exclusivity resulting from alterations in the same pathway(s).  This module contains five sets of genes (highlighted in orange in Figure~\ref{fig:real-data-results-subtypes}(b)) in the highest scoring collection $\mathbf{M}$ output by \OurAlgo{}. Consistent with TCGA study~\cite{TCGABRCA}, we find that: \textit{CDH1}, \textit{AKT1} and \textit{PIK3CA} are associated with the luminal A subtype, and these form a set in the \OurAlgo{} output.  Similarly, \textit{TP53} and amplification of chromosome region 4q13.3 are associated with the basal-like subtype, and also form a set in the \OurAlgo{} output. Two of the other sets contains genes in the same pathway.  \textit{PTEN} is a known inhibitor of \textit{PIK3CA}, explaining the observed exclusivity between \textit{PTEN} deletion and \textit{PIK3CA} mutation.  Moreover, \textit{MCL1}, \textit{MAP3K1}, \textit{AKT1} are all part of the PI(3)K/Akt signaling pathway.  Together, these sets contain five genes that are annotated as part of the PI(3)K/Akt signaling pathway in TCGA study~\cite{TCGABRCA}. (red circles in Figure~\ref{fig:real-data-results-subtypes}(b)).
% -- deletions of \textit{PTEN}, amplifications of \textit{MCL1}, mutations in \textit{MAP3K1}, \textit{AKT1}, and \textit{PIK3CA} -- are members of . This may implicate PI(3)K/Akt signaling as a key pathway in luminal A. 
%There are three additional mutually exclusive modules connected with three luminal A subtype genes, respectively. Two of them connect \textit{AKT1} and \textit{PIK3CA}, which show that there could be luminal A-specific sets of genes. 

The final set in this module includes mutations in the genes \textit{TP53}, \textit{CDH1}, \textit{GATA3} and \textit{CTCF}. These four genes are altered in $54.83\%$ ($278/507$) of the BRCA samples. \textit{TP53} is a member of the p53 signaling pathway, while \textit{CDH1}, \textit{GATA3}, and \textit{CTCF} all have been reported as potential driver genes in breast cancer. \textit{CDH1} is a tumor supressor that is well-known to play multiple roles in cancer \cite{Graziano2003}, including invasion and proliferation in breast cancer \cite{Hiraguri1998}. \textit{GATA3} is a transcription factor that has long been known to be involved in breast cancer tumorigenesis~\cite{usary2004gata3}. Recently, \textit{GATA3} has been reported to promote differentiation, suppresses metastasis and alter the tumor microenvironment in breast cancer~\cite{chou2013gata3}. As noted by Multi-Dendrix\cite{MultiDendrix}, \textit{GATA3} has also been reported to suppress tumor metastases through inhibition of \textit{CDH1} promoters \cite{Yan2010}, which suggests that the mutations in \textit{GATA3} are an alternate way to downregulate \textit{CDH1} and may explain the exclusivity of the mutations in \textit{GATA3} and \textit{CDH1}. 
Moreover, \textit{GATA3} is enriched for mutations in both luminal A and luminal B, i.e. 32 of the 54 mutations in \textit{GATA3} occur in luminal A ($P=0.0207$) and 32 of the 54 mutations in \textit{GATA3} occur in luminal B ($P=0.065$). This might suggest \textit{GATA3} mutations mainly occur in patients with luminal breast cancer.
\textit{CTCF} neighbors \textit{CDH1} on chromosome 16q22.1 and has been reported with \textit{CDH1} to be a tumor suppressor in breast cancer \cite{cowin2005cdh1,green2009ctcf}. Interestingly, both \textit{CDH1} and \textit{CTCF} have most of their mutations in samples of the luminal A subtype. \textit{CDH1} is enriched for mutations in luminal A (as reported in~\cite{TCGABRCA}) and 9 of the 13 mutations in \textit{CTCF} occur in luminal A ($P=0.0891$), suggesting these two genes are in a pathway specifically targeted in luminal A. Furthermore, 4 of the 9 mutations in \textit{CTCF} in luminal A are missense mutations in zinc finger domains, suggesting possible functional role for these mutations \cite{filippova2002tumor}.
 
Together, these results demonstrate \OurAlgo{}'s ability to  simultaneously identify alterations that are mutually exclusive due to interactions between genes in pathways or due to subtype-specific alterations.  This allows a more refined interpretation of mutually exclusive alterations than simple pairwise analyses.

\section{Discussion}
\label{sec:discussion}
We introduce the \OurAlgo{} algorithm for identifying collections of mutually exclusive alterations in cancer \emph{de novo}, i.e. with no prior biological knowledge. \OurAlgo{} uses a novel statistical score for exclusive alterations that conditions on the frequency of each alteration and thus can detect exclusivity of rare mutations. \OurAlgo{} overcomes large computational challenges in computing the score using a new algorithm for contingency table analysis, and in optimizing the score in genome-scale data using the first Markov chain Monte Carlo (MCMC) algorithm for identifying collections of exclusive alterations. 

We demonstrate that \OurAlgo{} is superior to earlier \emph{de novo} methods -- Dendrix~\cite{Dendrix}, muex~\cite{muex}, and Multi-Dendrix~\cite{MultiDendrix} -- on simulated and real data. We then apply \OurAlgo{} to large mutation datasets from multiple TCGA cancer types \cite{TCGAGBM2008,TCGAAML,TCGABRCA,TCGASTAD}. On each dataset, \OurAlgo{} identifies significantly exclusive collections of alterations that overlap well-known cancer pathways, as well as implicate novel cancer genes. In addition, \OurAlgo{} illustrates subtle relationships between mutual exclusivity resulting from cancer subtypes and exclusivity resulting from pathways or protein interactions.  These findings provide testable hypotheses for further downstream analysis or experimental validation.

The input to \OurAlgo{} is a matrix of binary alterations, and thus can be used to analyze a variety of alterations including point mutations and indels, copy number aberrations (amplifications and deletions) and complex rearrangements, splice-site mutations, gene fusions, and subtype annotations. \OurAlgo{} may be useful in analysis of other types of alterations; e.g. germline variants. 

Another application for \OurAlgo{} is pan-cancer analysis, such as the recently published TCGA study \cite{TCGAPanCan} and the upcoming ICGC pan-cancer project. Since pan-cancer datasets have many cancer-type specific alterations, \OurAlgo{}'s ability to simultaneously analyze type-specific and other types of exclusive alterations should prove useful for this analysis.
Finally, we anticipate that the novel tail enumeration strategy used in  \OurAlgo{} may be of broader interest, both for examining mutual exclusivity in other datasets, including non-biological data, as well as for adapting for other types of exact statistics.

\clearpage
\bibliographystyle{unsrt} % Style BST file
\bibliography{references}      % Bibliography file (usually '*.bib' )

\clearpage
\vspace{0.15in}
\begin{minipage}{0.95\linewidth}
\begin{center}
{\Large
	Supplementary Information for \\
	\vspace{0.1in}
	\textbf{\OurAlgo{}: A Statistical Approach to Identify Combinations of Mutually Exclusive Alterations in Cancer}
} \\
\vspace{0.15in}
{\normalsize
Mark D.M. Leiserson$^{1,2,*}$, Hsin-Ta Wu$^{1,2,*}$, Fabio Vandin$^{1,2,3}$, Benjamin J. Raphael$^{1,2}$\\
}

\vspace{0.25in}

{\small 
\emph{$^1$Department of Computer Science and $^2$Center for Computational Molecular Biology, Brown University, Providence, RI, USA}\\
\emph{$^3$Department of Mathematics and Computer Science, University of Southern Denmark, Odense M, Denmark}\\
\emph{$^{*}$Equal contribution.}\\
	\vspace{0.1in}
	Correspondence:
	\texttt{braphael@brown.edu}
}
\end{center}
\end{minipage}
\vspace{0.25in}
% Reset Figure/Table/Sections counters and prepend S to them
\setcounter{figure}{0}
\makeatletter 
\renewcommand{\thefigure}{S\@arabic\c@figure} 

\setcounter{table}{0}
\makeatletter 
\renewcommand{\thetable}{S\@arabic\c@table} 

\setcounter{section}{0}
\makeatletter 
\renewcommand{\thesection}{S\@arabic\c@section} 

\section{Methods}
\subsection{MCMC Algorithm}
\label{sec:mcmc-alg}
We define a Markov chain whose states $\Omega$ are possible collections $\mathbf{M}$ and where transitions between states (collections) are defined such that the chain is ergodic. Finite and ergodic Markov chains converge to a unique stationary distribution. In this case, because we want to sample from collections $\mathbf{M}$ in proportion to their weights $\Phi(\mathbf{M})^{-1}$, our desired stationary distribution is 
\begin{equation}
\pi_{\mathbf{M}} = \frac{\Phi(\mathbf{M})^{-1}}{\sum_{\mathbf{M}' \in \Omega} \Phi(\mathbf{M'})^{-1}}.
\end{equation}
Note that we use $\Phi(\mathbf{M})^{-1}$ so more exclusive collections have higher weights. The Metropolis-Hastings algorithm \cite{Metropolis1953,Hastings1970} is a method for defining transition probabilities for an irreducible Markov chain such that the modified chain is ergodic and has a desired stationary distribution. A Metropolis-Hastings algorithm to sample collections $\mathbf{M}$ according to this stationary distribution is as follows:
\\
It is easy to see that this chain is ergodic (it is possible to reach any state (collection) from any other state (collection), it is finite, and it is not bipartite) and thus it converges to our desired stationary distribution. We apply a parameter $\alpha$ to increase/decrease the difference between $\Phi(\mathbf{M}'_N)$ and $\Phi(\mathbf{M}_N)$. Also, in the second step of the algorithm, we ensure that the number of exclusive alterations is larger than the number of co-occurring by checking that the Dendrix weight $W(M) > 0$. This is to avoid examining sets alterations with high coverage (e.g. altered over $90\%$ of samples) that may have significant exclusivity even though relatively few samples harbor exclusive alterations. We assess convergence of the MCMC algorithm by calculating \emph{total variation distance} of the the sampling distributions from multiple chains with different initializations (See Supplementary Section \ref{sec:mcmc-convergence}). The MCMC algorithm consists of the following steps:

\begin{description}
\item[Initialization.] Choose $tk$ genes uniformly at random from $\mathcal{E}$, and assign $k$ genes at random to initialize $\mathbf{M} = M_1, \dots, M_t$.
\item[Iteration.] For $N=1, 2, \dots$, obtain $\mathbf{M}_{N+1}$ from $\mathbf{M}_N$ as follows:
	\begin{enumerate}
	\item Select a gene $g$ uniformly at random from $\mathcal{E}$.
	% TODO: use enumerate below (compile error with GenomeBiology for some reason)
	\item Define the proposed collection $\mathbf{M}'_N$ as follows:
		\\
		i) If $g \notin \mathbf{M}_N$, then choose uniformly at random gene $g' \in M_i$, and replace $g'$ with $g$.
		\\
		ii) Else, choose uniformly at random gene $g' \in M_i$, and \emph{swap} genes $g$ and $g'$. Note that if $g, g' \in M_i$, then $M_i$ will be unchanged.
	\item Let $P(\mathbf{M}_N, \mathbf{M}'_N) = \min\{ 1, \frac{\Phi(\mathbf{M}_N)^{\alpha}}{\Phi(\mathbf{M}'_N)^{\alpha}}\}.$
	\item With probability $P(\mathbf{M}_N, \mathbf{M}'_N), \mathbf{M}_{N+1} = \mathbf{M}'_N$, else
		$\mathbf{M}_{N+1} = \mathbf{M}_N$.
	\end{enumerate}
\end{description}

\paragraph{Convergence of MCMC from different initial gene sets}
\label{sec:mcmc-convergence}
To assess convergence of the MCMC algorithm, we compare the sampling distributions from multiple chains with different initializations.  The idea is that if multiple chains have converged, by definition they should appear very similar to one another; if not, one or more of the chains has failed to converge. We create the following pipeline for performing \OurAlgo{} on all our experiments. To make the initializations have high variety, we create $5$ to $10$ initializations for any $t$ and $k$ of \OurAlgo{}. One initialization is from Multi-Dendrix~\cite{MultiDendrix} results with the same $t$ and $k$, and other initializations generate randomly. 

Precisely, we first define whole sampling gene sets $\Omega$ as the union of the last $0.5\%$ of sampling gene sets $\omega_i$ of each chain $i$. We then define the distribution for each chain $i$ over the whole sampling gene sets $\Omega=\cup_{\forall i} \omega_i$ as $P_i(s) =  F_i(s) / |\omega_i|$ if $s \in \Omega $ othereise $0$, and the distribution for the union chain $u$ over the whole sampling gene sets $\Omega$ as $P_u(s) =  F_u(s) / |\Omega|$. We calculate the mean total variation distance over $P_i$ and $P_u$. The total variation distance between $P_i$ and $P_u$ is defined as

\begin{equation}
|P_i -  P_u ||_{TV} = \max_{s \in \Omega } \|P_i(s) - P_u(s)\|
\end{equation}

We start \OurAlgo{} with $100$ million iterations for each of these initializations. We examine convergence by calculating both metrics across the initializations. If the metric is close to $0$, this indicates the convergence and the process will be stopped; otherwise, we increase $1.5$ times of the number of iterations until the number of iterations touches $1$ billion. In final, we use the union of ten sampling distribution as sampling results.  For example, we performed above pipeline for $t=3$ and $k=4$ on AML mutation data and plotted the distribution of total variation distance after $1$ million iteration and $10$ million iteration MCMC runs (Figure ~\ref{fig:aml_k444_exact_1Mvs10M_total_variation_distance}). 

\subsection{Parameter selection}
% of parameter $\delta$ in marginal probability graph}
\label{sec:parameter-selection}
We select $\delta$ with the following heuristic procedure. When we run \OurAlgo{} with $t$ sets in the collection, ideally we should obtain $t$ cliques in the marginal probability graph. To find the best $\delta$ that fulfills the expectation, we search for an ``L-corner'' in a graph of the number of edges in the marginal probability graph as a function of the edge weight.

% scan edge weights in descending order and choose $\delta$ at the first ``L-corner'' of two best-fit lines and make sure the number of edges in the subgraph is at least $t \times {k \choose 2}$.

More precisely, we first plot a log-log distribution with the number of edges in the marginal probability graph with edge weight $\ge p$ against edge weight $p$ (Figure~\ref{fig:delta_selection}). We choose $\delta$ starting from the minimum edge weight $p_{min}$ that contains at least $t \times {k \choose 2}$ edges in the marginal probability graph. e.g. the yellow horizontal line in Figure~\ref{fig:delta_selection} shows the number of edges 
%$9$ 
in GBM with $k=3$ and $t=3$. We identify a value $\delta$ where the number of edges increases dramatically after this value as the probability threshold decreases. To find this value, for each value $x$ we perform a linear regression of two best-fit lines (using root mean squared error) before and after this value.  We the first $p > p_{min}$ that forms a ``L-corner'', i.e. the slope of the two best-fit lines changes from  a smaller negative value to a larger negative value as the value $x$ decreases (e.g. moving leftward in Figure~\ref{fig:delta_selection}).

For each cancer dataset, we ran \OurAlgo{} with $k=4$, $t=4$, and $100$ million iterations using $5$ to $10$ random initializations. For BRCA and STAD with subtypes, we ran \OurAlgo{} with $k=4$ and $t$ equal to the number of pre-defined subtypes (4 and 3, respectively), and $100$ million iterations using $10$ random initializations. (See Section \ref{appendix:mutation_data} and Supplementary Section \ref{sec:mcmc-convergence}). Ideally, \OurAlgo{} should be run with the largest values of $k$ and $t$ that are biologically meaningful for a particular dataset.  If smaller values of $k$ and $t$ are best supported by the data, the summarization procedure will demonstrate this.  In practice, using large values of $k$ and $t$ might lead to long run times and slow convergence of the MCMC algorithm, since the space of possible collections will be very large.  Thus, an alternative approach that we use to generate results is to run with small values of $t$ and $k$ (e.g. $t = 3,4$ and $k = 3,4$) and examine the resulting marginal probability graph.  If there $t$ or more cliques or approximate cliques in the graph, this suggests the use of larger values of $t$ and $k$. We used this approach to find larger collections in the AML dataset (See details in Section \ref{sec:real-data}).

\section{Data}

\subsection{Somatic mutation datasets}
\label{appendix:mutation_data}
\paragraph{Acute myeloid leukemia (AML)}
\label{sec:data-aml}
The AML dataset contains whole-exome and copy number array data in $200$ AML patients from The Cancer Genome Atlas (TCGA) ~\cite{TCGAAML}. Using the annotations in \cite{TCGAAML}, we categorized multiple genes together based on expert knowledge, which results in 9 categories including spliceosome, cohesin complex, MLL-X fusions, other myeloid transcription factors, other epigenetic modifiers, other tyrosine kinase, serine/threonine kinase, protein tyrosine phosphatase, and RAS protein.  More details are in ~\cite{TCGAAML}. This results in $51$ genes and $200$ patients.

\paragraph{Glioblastoma multiforme (GBM)}
\label{sec:data-gbm}
We analyzed three GBM datasets:
\begin{enumerate}
\item The Multi-Dendrix GBM dataset from \cite{MultiDendrix}. This dataset contains whole-exome and copy number array data in $261$ GBM patients and $398$ genes from The Cancer Genome Atlas (TCGA) ~\cite{TCGAGBM2008}. Data preparation for GBM can be found in ~\cite{MultiDendrix}. Note that in Section \ref{sec:gbm} we included amplifications in \textit{EGFR} which were not considered in  \cite{MultiDendrix}.
\item The muex GBM dataset from \cite{muex}. This dataset contains 83 alterations in 236 samples from~\cite{TCGAGBM2008}, including SNVs in genes identified as significantly mutated by MutSigCV~\cite{Lawrence2013} and CNAs called by GISTIC2~\cite{Mermel2011} then restricted to those with significantly concordant gene expression (higher for amplifications, lower for deletions).
\item The Pan-cancer GBM dataset from \cite{TCGAPanCan}. We analyzed the non-silent mutations (single nucleotide variants and small indels) from the MAF file and focal copy number aberrations from GISTIC2 output. This dataset contains 509 genes in 291 samples. Moreover, we removed genes with non-silent mutations in $<1\%$ of samples and with mutations in $>2.5\%$ of samples with MutSigCV \cite{Lawrence2013} \textit{q}-value $>0.1$. This dataset contains 406 genes in 291 samples.
\end{enumerate}

\paragraph{Gastric cancer (STAD)}
\label{sec:data-stad}
We analyzed the non-silent mutations (single nucleotide variants and small indels) from the MAF file in 289 gastric cancer samples. We also included focal driver copy number aberrations from GISTIC2 output via Firehose, fusion genes, rearrangements and splicing events~\cite{TCGASTAD}. We removed 74 hypermutators and genes with non-silent mutations in $<2.5\%$ of samples and with mutations in $>3\%$ of samples with MutSigCV \cite{Lawrence2013} \textit{q}-value $>0.25$. This process results in $217$ STAD patients and $397$ genes with mutations. We considered four subtypes identified by TCGA~\cite{TCGASTAD}, including tumors were positive for Epstein-Barr virus (EBV), tumors had high microsatellite instability (MSI)
% which is the tendency for mutations to accumulate in repeated sequences of DNA, 
genomically stable (GS) tumors with a low level of somatic copy number aberrations, and chromosomally unstable (CIN) tumors with a high level of somatic copy number aberrations and were called.  We do not analyze the MSI subtype since samples in MSI are hypermutators.

\paragraph{Breast cancer (BRCA)}
\label{sec:data-brca}
The BRCA dataset contains whole-exome and copy number array data in $507$ BRCA patients and $375$ genes from The Cancer Genome Atlas (TCGA) ~\cite{TCGABRCA}. Data preparation for BRCA can be found in ~\cite{MultiDendrix}. We downloaded subtype information of BRCA from The Cancer Genome Atlas (TCGA) ~\cite{TCGABRCA}. We considered four subtypes -- basal-like, HER2-enriched, luminal A, and luminal B -- that each contain at least $10\%$ of the total samples.

\subsection{Simulated data}
\label{sec:simulated-data}
We generated simulated datasets using the following approach. Recall $\mathbf{C}$ is a set of highly altered genes whose alterations are not necessarily exclusive.
\begin{packed_enum}
\item Select $k$ genes to form an ``implanted pathway'' $P$.
\item Let $\gamma_P$ be the fraction of mutated samples in $P$. Select $\gamma_P \times n$ samples to be
	exclusively mutated in $P$, where the proportion of mutations in each gene in $P$ is given by the
	tuple $\mu_P=(c_1, \dots, c_k)$.
\item Randomly select samples to be mutated in each gene in $\mathbf{C}$, where the fraction of mutated
	samples per  gene is given by $\gamma_\mathbf{C}$.
\item For each of the $n$ samples $s$ in each of the $m$ genes $g$ (including the implanted and cancer genes),
	mutate $g$ in $s$ with fixed probability $q$. This step introduces noise into the dataset.
\end{packed_enum}
We used $m=100$, $n=500$, $k=3$, $\mu_P=(0.5, 0.35, 0.15)$, $|\mathbf{C}|=5$, $\gamma_\mathbf{C}=(0.67, 0.49, 0.29, 0.29, 0.2)$, and $q=0.0027538462$.\footnote{We chose values for $\mathbf{C}$ and $q$ using values calculated from real data. We choose $C$ to match the mutation frequencies of the five most mutated genes in the TCGA glioblastoma dataset. We calculated $q$ empirically from the TCGA breast cancer mutation matrix.} We removed alterations that occurred in fewer than $5$ alterations (resulting in the average number of genes of $276.44$). We ran \OurAlgo{} $100$ million iterations from $3$ random initial starts.

\section{Supplementary Tables}
\FloatBarrier
\setcounter{section}{0}
\makeatletter 
\renewcommand{\thesection}{S\@arabic\c@section} 

\setcounter{figure}{0}
\makeatletter 
\renewcommand{\thefigure}{S\@arabic\c@figure} 

\setcounter{table}{0}
\makeatletter 
\renewcommand{\thetable}{S\@arabic\c@table}

\begin{table*}[h!]
\footnotesize
\centering
\tabcolsep=0.08cm
\begin{tabular}{ @{}llrllr@{} }
\toprule
\multicolumn{3}{c}{Multi-Dendrix} & \multicolumn{3}{c}{\OurAlgo{}} \\ 
\cmidrule(l){1-3}\cmidrule(l){4-6}
Alterations set & $\Phi(M)$ & W(M) & Alteration set & $\Phi(M)$ & W(M) \\ 
\cmidrule(l){1-6} 
\multicolumn{6}{c}{Pan-cancer BRCA dataset \cite{leiserson2014pan}, $k=4$, $t=4$}\\
\cmidrule(l){1-6} 
\textit{PIK3CA}, \textit{MCL1}(A), \textit{ZNF703}(A), \textit{AKT1} & $9.2\times 10^{-9}$ & 403 &
\textit{PIK3CA}, \textit{ING5}(D), \textit{PEG3}, \textit{AKT1} & $2.6\times10^{-12}$ & 352  \\
\textit{TP53}, \textit{GATA3}, \textit{MAP3K1}, \textit{CDH1} & $1.7\times 10^{-11}$ & 397 &
\textit{TP53}, \textit{GATA3}, \textit{CDH1}, \textit{CTCF}  & $4.1\times 10^{-17}$ &  380 \\
\textit{CCND1}(A), \textit{MYC}(A), \textit{MAP2K4}, \textbf{\textit{CBFB}} & $5.4\times 10^{-1}$  & 307 & 
\textit{MT-ND1}, \textit{MYC}(A), \textit{LAMA2}, \textit{MAP3K1} & $1.7\times 10^{-7}$  & 269 \\
\textit{TUBD1}(A), \textbf{\textit{STK11,TCF3}(D)}, \textbf{\textit{MLL3}}, \textbf{\textit{RUNX1}} &  $2.2\times 10^{-1}$  & 284 & 
\textbf{\textit{CBFB}}, \textbf{\textit{FHIT}(D)}, \textbf{\textit{RUNX1}}, \textbf{\textit{ZNF703}(A)} & $5.5\times10^{-3}$ & 256 \\
 \cmidrule(l){1-6} 
\multicolumn{6}{c}{Pan-cancer BRCA dataset \cite{leiserson2014pan} without MutSigCV filter, $k=4$, $t=4$}\\
\cmidrule(l){1-6} 
\textit{PIK3CA}, \textit{TUBD1}(A), \textbf{\textit{PTEN}(D)}, \textit{AKT1} & $2.0\times 10^{-7}$ & 397 &
\textit{PIK3CA}, \textit{ING5}(D), \textit{PEG3}, \textit{AKT1} & $2.6\times10^{-12}$ & 352  \\
\textit{TP53}, \textit{GATA3}, \textit{CDH1}, \textit{MAP2K4} & $5.0\times 10^{-16}$ & 382 &
\textit{TP53}, \textit{GATA3}, \textit{CDH1}, \textit{CTCF}  & $4.1\times 10^{-17}$ &  380 \\
\textit{CCND1}(A), \textit{MYC}(A), \textbf{\textit{MUC4}}, \textit{MAP3K1} & $2.7\times 10^{-1}$  & 323 & 
\textit{MT-ND1}, \textit{MYC}(A), \textit{LAMA2}, \textit{MAP3K1} & $1.7\times 10^{-7}$  & 269 \\
\textit{MCL1}(A), \textit{ZNF703}(A), \textbf{\textit{ENSG1*}}, \textbf{\textit{CROCCP2}} & $1.2\times 10^{-2}$  & 294 & 
%\textit{CBFB}, \textit{FHIT}(D), \textit{RUNX1}, \textit{ZNF703}(A) & $4.4\times10^{-3}$ & 256 \\
\textbf{\textit{ERBB2}(A)}, \textbf{\textit{PTEN}(D)}, \textbf{\text{CROCCP2}}, \textbf{\textit{ENSG2*}}&  $4.4\times 10^{-5}$ & 249 \\
 \bottomrule
\end{tabular}
\caption{\textbf{Collections of alterations reported by Multi-Dendrix and \OurAlgo{} on TCGA Pan-cancer breast cancer data \cite{leiserson2014pan} with $k=3$ and $t=4$.} Bolded alterations indicate differences between alteration datasets with and without MutSigCV filter. *Due to the limited width of the page, we replaced \textit{ENSG00000210082} as \textit{ENSG1} and \textit{ENSG00000211459} as \textit{ENSG2}. }
\label{tbl:mdendrix_comp_brca}
\end{table*}

\begin{table*}[h!]
\begin{center}
\begin{tabular}{cccc}
\toprule
& $\mathbf{t=2}$ & $\mathbf{t=3}$ & $\mathbf{t=4}$\\
\hline
Coverage & (0.70, 0.50) & (0.70, 0.60, 0.50) & (0.70, 0.60, 0.50, 0.40)\\
\bottomrule
\end{tabular}
\end{center}
\caption{Coverages $\gamma_\mathbf{P}$ for each simulated dataset with non-overlapping pathways.}
\label{tbl:sim-coverages}
\end{table*}%

\begin{table*}[h!]
\begin{center}
\begin{tabular}{cclccccc}
\toprule
& & & \multicolumn{3}{c}{\textit{\OurAlgo{}}} & \multicolumn{2}{c}{\textit{Multi-Dendrix}}\\
$\mathbf{t}$ & $\mathbf{k}$ & \textbf{Class} & \textbf{Consensus} & \textbf{True} & \textbf{Highest weight} & \textbf{Consensus} & \textbf{True} \\
\hline
2 & 3 & Non-overlapping & 0.53 & 1.0 & 0.65 & 0.43 & 0.49\\
2 & 4 & Non-overlapping & 0.51 & 1.0 & 0.79 & 0.53 & 0.48\\
2 & 5 & Non-overlapping & 0.74 & 1.0 & 0.76 & 0.45 & 0.48\\
3 & 3 & Non-overlapping & 0.68 & 0.97 & 0.85 & 0.45 & 0.65\\
 &  & Overlapping & 0.66 & 0.94 & 0.78 & 0.48 & 0.69\\
3 & 4 & Non-overlapping & 0.88 & 1.0 & 1.0 & 0.51 & 0.65\\
 &  & Overlapping & 0.82 & 0.95 & 0.95 & 0.69 & 0.68\\
3 & 5 & Non-overlapping & 0.94 & 1.0 & 0.88 & 0.57 & 0.64\\
 &  & Overlapping & 0.97 & 0.96 & 0.91 & 0.58 & 0.6\\
4 & 3 & Non-overlapping & 0.8 & 0.97 & 0.73 & 0.37 & 0.5\\
4 & 4 & Non-overlapping & 0.92 & 1.0 & 0.85 & 0.47 & 0.54\\
4 & 5 & Non-overlapping & 0.83 & 1.0 & 0.73 & 0.47 & 0.5\\
\bottomrule
\end{tabular}
\end{center}
\caption{\textbf{Results of \OurAlgo{} and Multi-Dendrix on simulated datasets consisting of $t$ implanted pathways, each with $k$ genes.} The mean average adjusted Rand index across $25$ simulated datasets for different parameter choices. We compared \OurAlgo{} to Multi-Dendrix using each algorithm's consensus procedure (``Consensus''). We also compared \OurAlgo{} to Multi-Dendrix when run with the true values of $t$ and $k$ of the implanted pathways (``True``).}
\label{tbl:collection-sims}
\end{table*}%
\clearpage

\section{Supplementary Figures}
\FloatBarrier
\begin{figure*}[h!]
\centering
	\includegraphics[width=\textwidth]{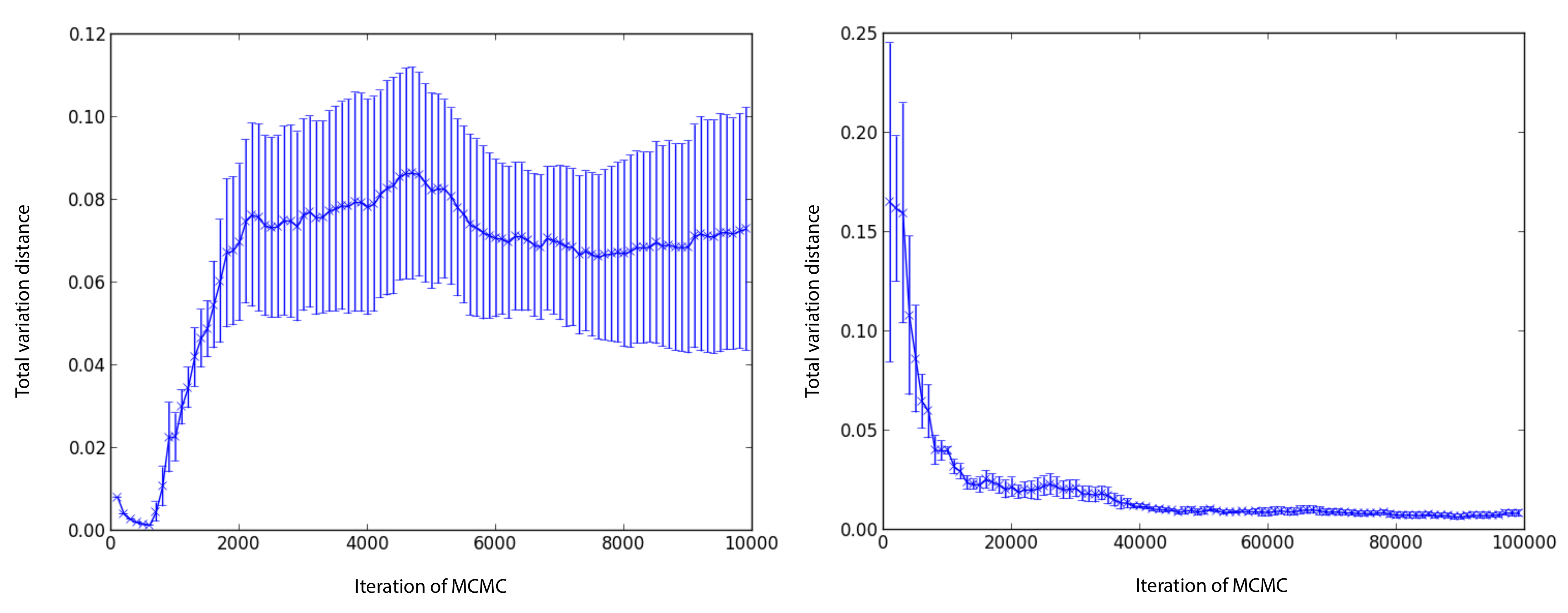}
	\caption{\textbf{Plots of total variation distance distribution in each iteration of for an MCMC run with $1M$ iterations (left) and an MCMC run with $10M$ iterations (right).}}
	\label{fig:aml_k444_exact_1Mvs10M_total_variation_distance}
\end{figure*}

\begin{figure*}[h!]
\begin{center}
\includegraphics[width=0.6\textwidth]{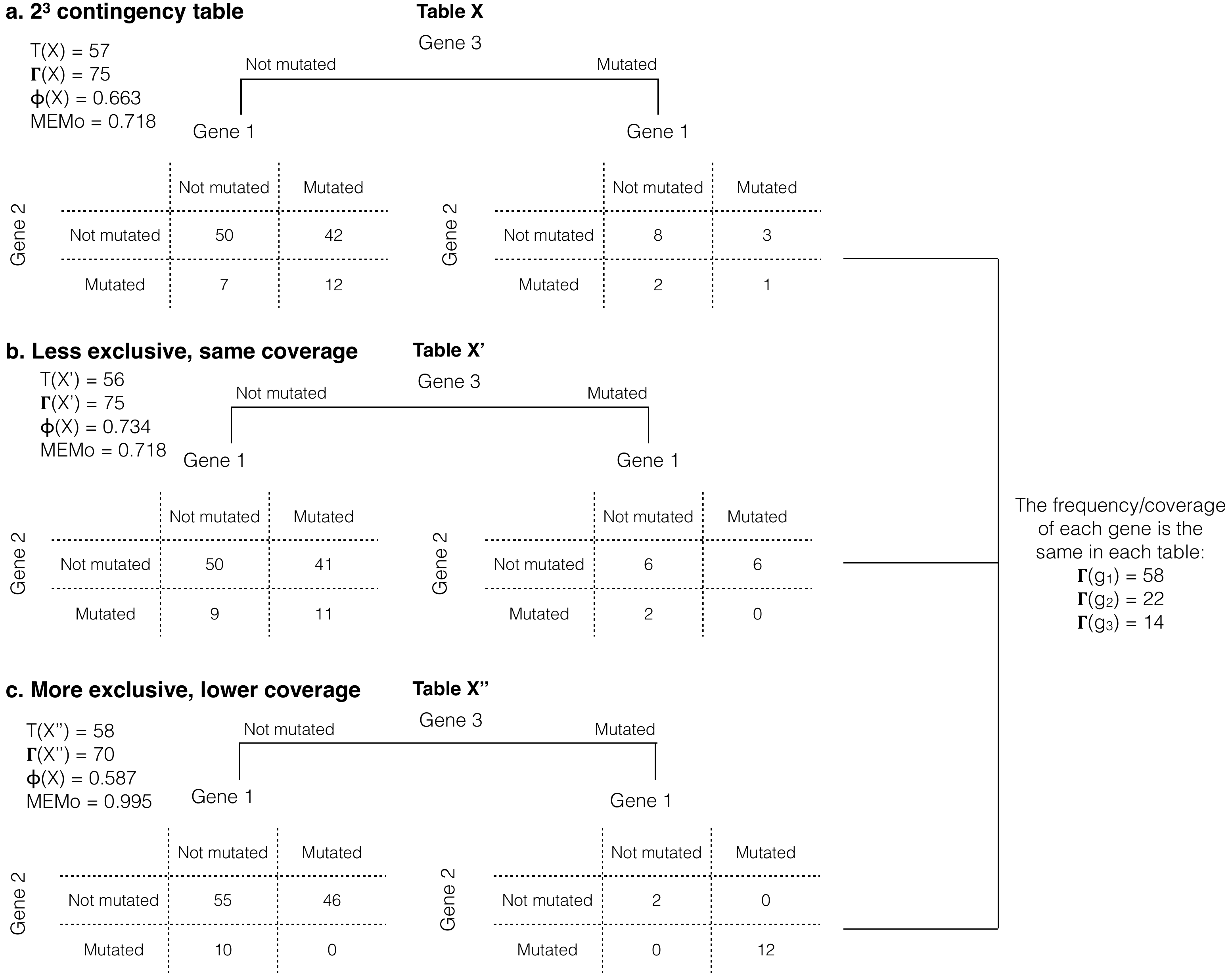}
\caption{\textbf{Two cases where the MEMo permutation test statistic $\Gamma$ (the coverage, or number of altered samples) deflates or inflates the $P$-value compared to the \OurAlgo{} test statistic $T$ (the number of samples with exclusive mutations).}
}
\label{fig:memo-comparison}
\end{center}
\end{figure*}

\begin{figure*}[h!]%{1\textwidth}
\begin{center}
  \includegraphics[width=1\textwidth]{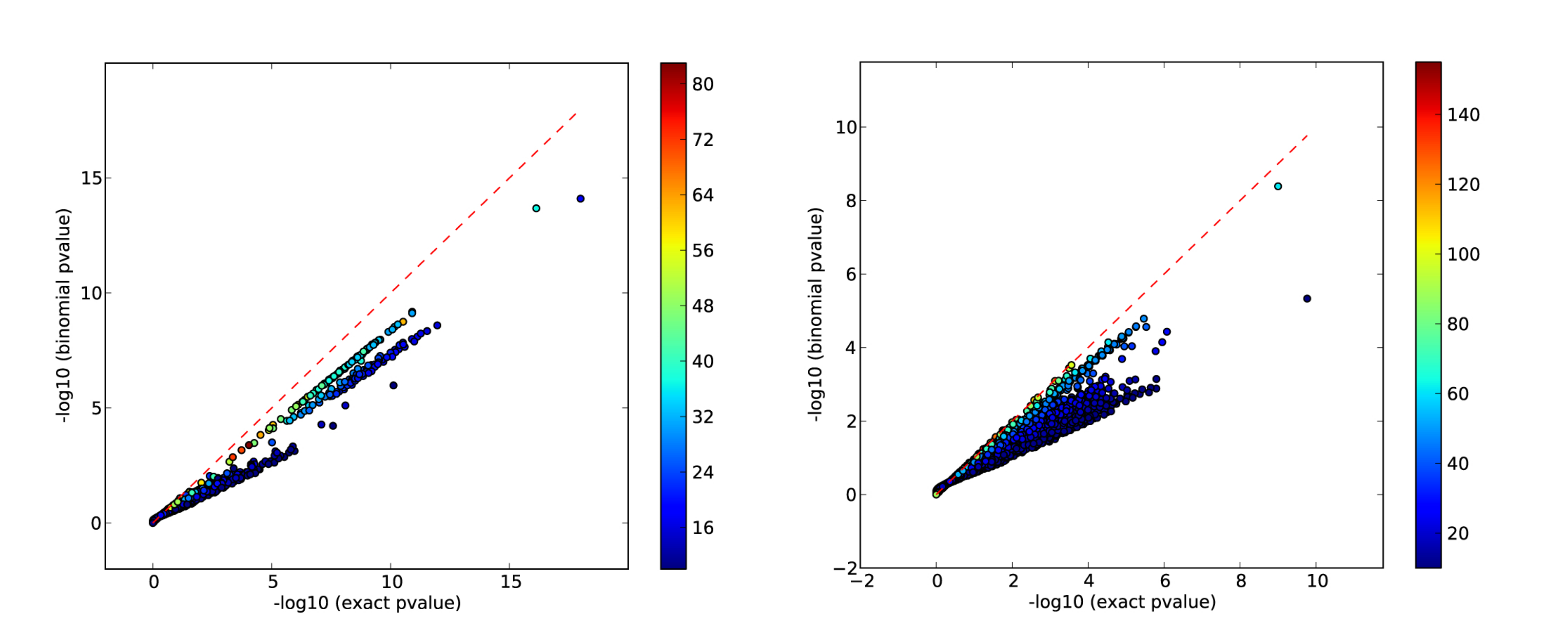}
  \caption{\textbf{Scatter plot between negative log of exact and binomial $P$-values for all sets of $k=3$ alterations on the GBM dataset (left) and BRCA dataset (right).} The color of each dot represents the number of co-occurring alterations according to the scale at the right. Note that the $P$-values for the exact test much smaller than the binomial only in cases with relatively low number of co-occurrences.  These cases are the fastest to compute with the tail enumeration algorithm.}
  \label{fig:approx_binom_exact_vs_cooccurring}
\end{center}
\end{figure*}

\begin{figure*}[h!]%{1\textwidth}
\begin{center}
 \includegraphics[width=0.9\textwidth]{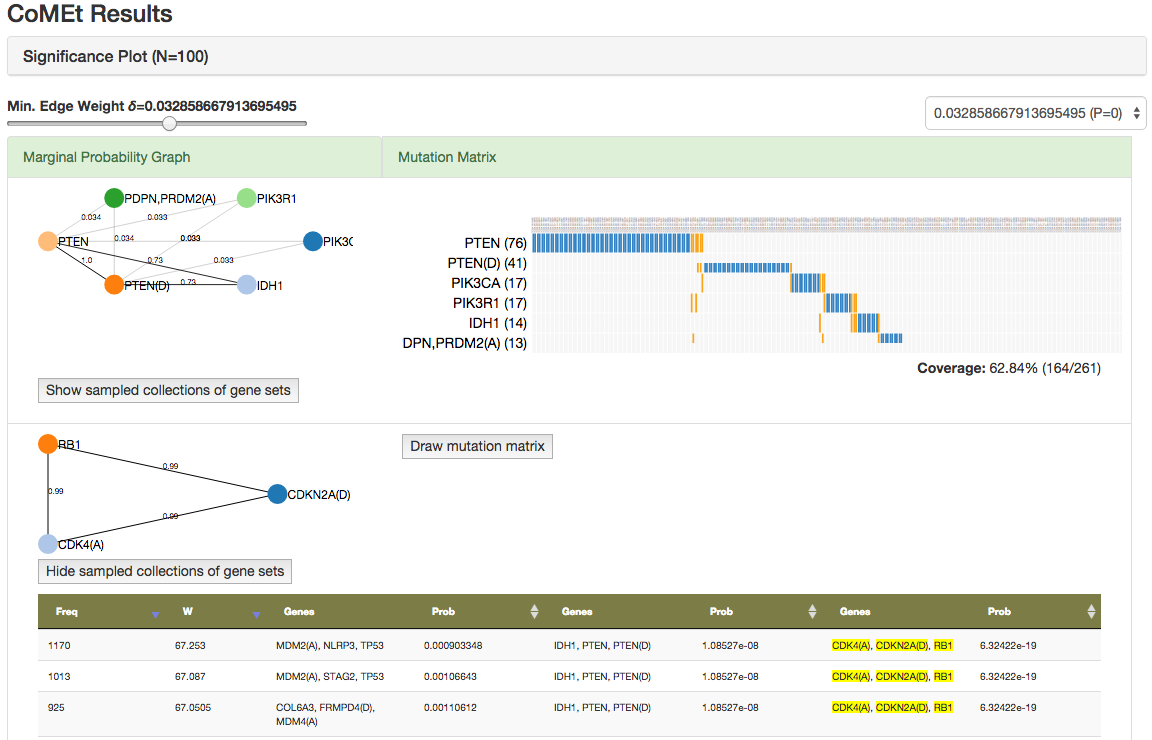}
 \caption{\textbf{Screenshot of the web application for interactive visualization of \OurAlgo{} results.}}
 \label{fig:website}
\end{center}
\end{figure*}

\begin{figure*}[h!]%{1\textwidth}
\begin{center}
 \includegraphics[width=0.8\textwidth]{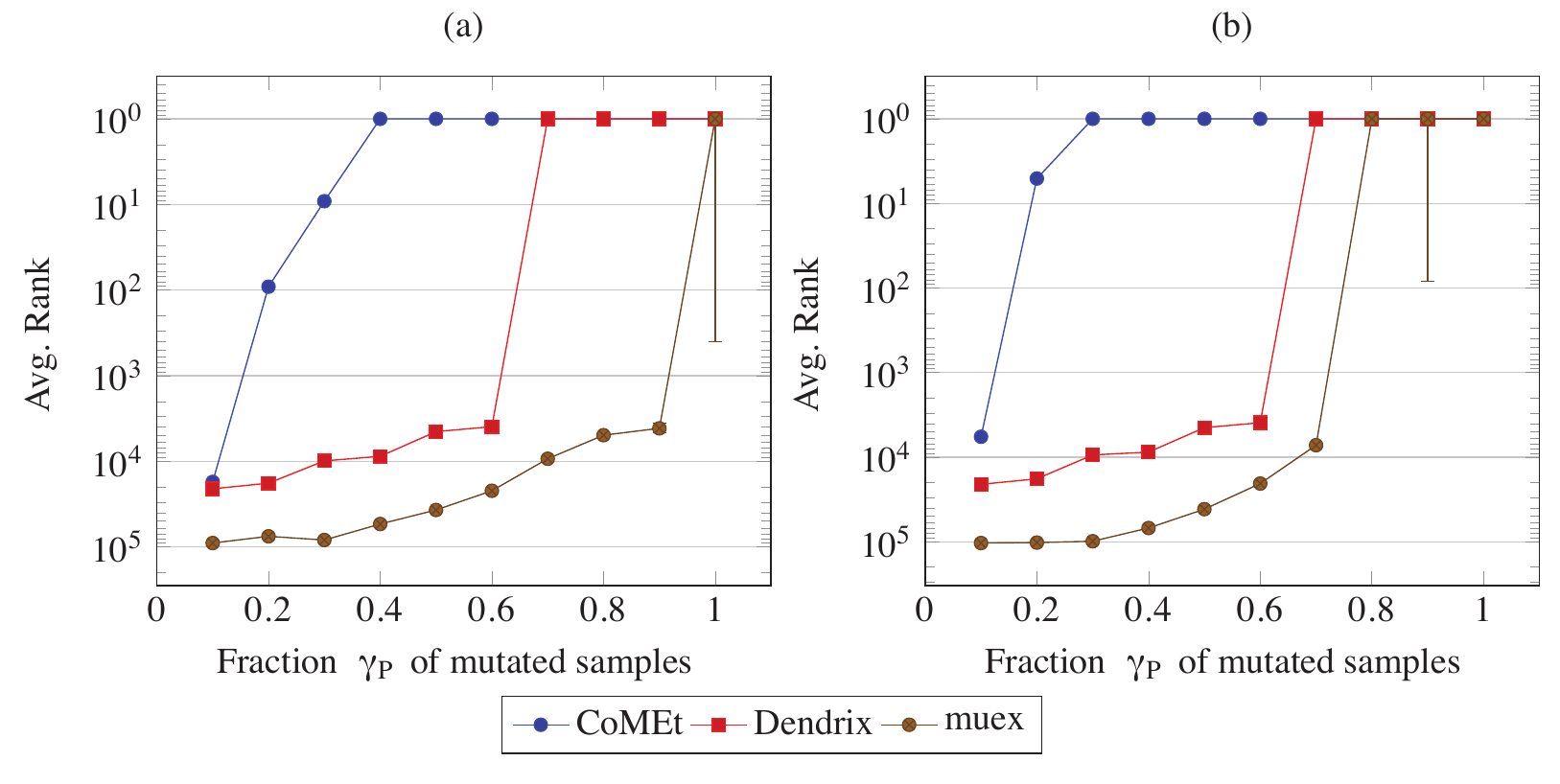}
\caption{\textbf{The average rank of the implanted pathway in output from \OurAlgo{} (blue), Dendrix (red), and muex (green) in 25 simulated datasets using (a) $n=250$ and (b) $n=750$ samples.} The average number of ties (i.e. gene sets with the same score as the implanted pathway) are shown as error bars.}
\label{fig:varying-size-sims}
\end{center}
\end{figure*}

\begin{figure*}[h!]%{1\textwidth}
\begin{center}
 \includegraphics[width=0.6\textwidth]{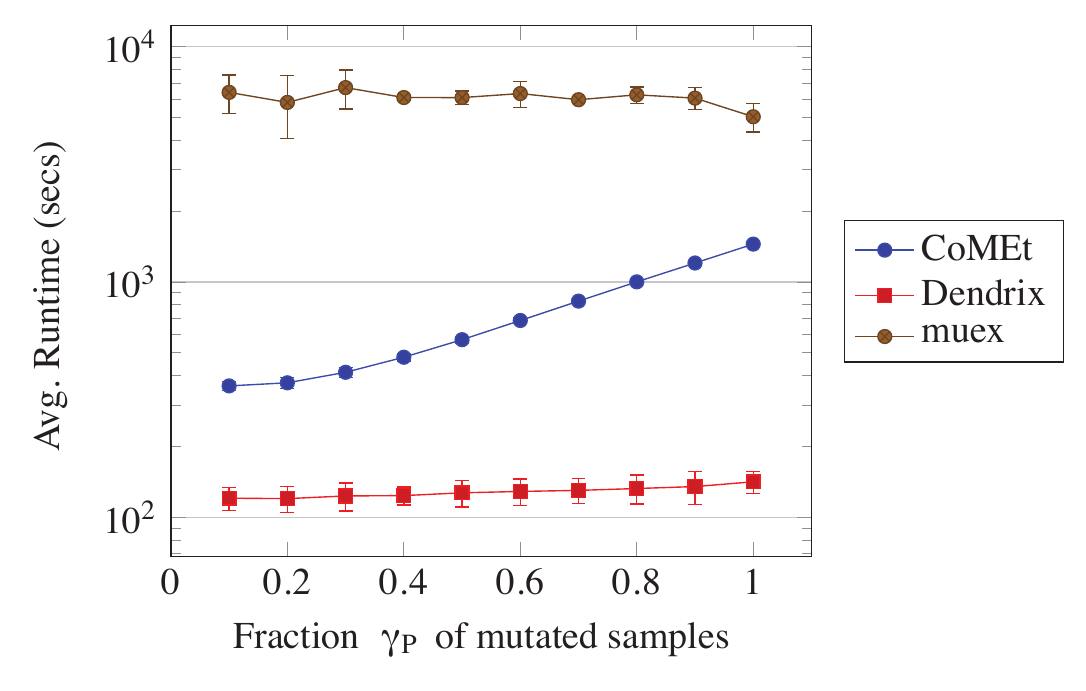}
\caption{\textbf{Comparison of the average runtime of  \OurAlgo{}, Dendrix, and muex on $N=25$ simulated datasets with a single implanted pathway.} The standard deviation for the runtime is shown as an error bar.}
\label{fig:sim-runtimes}
\end{center}
\end{figure*}

\begin{figure*}[h!]
\begin{center}
\includegraphics[width=0.5\textwidth]{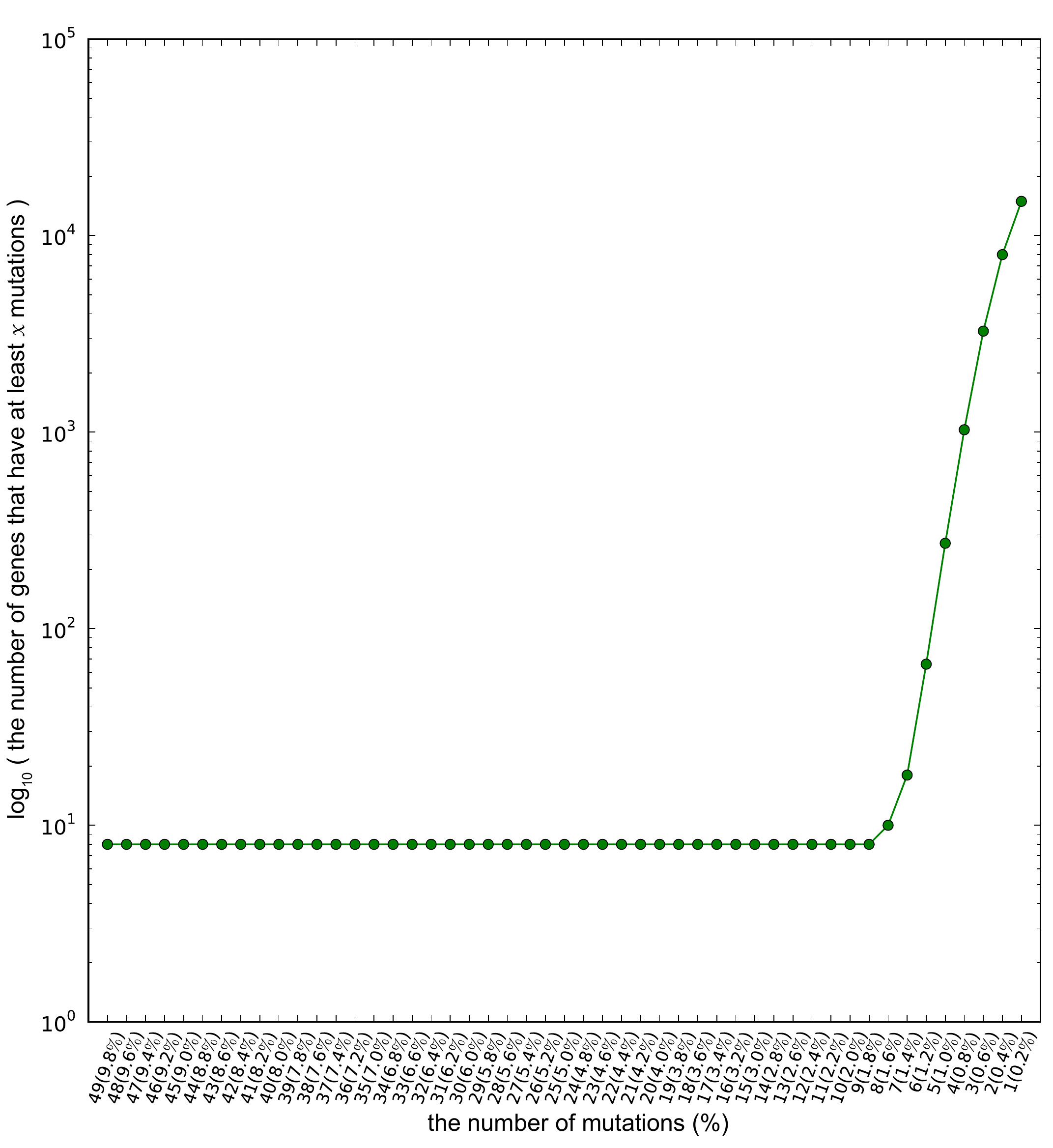}
\caption{\textbf{The distribution of the number of genes with  $ \ge x$ mutations in simulated data.} We removed those genes mutated in fewer than $1\%$ of mutations, i.e. genes mutated in fewer than 5 samples.}
\label{fig:simulation_cutoff}
\end{center}
\end{figure*}

\begin{figure*}[h!]
\begin{center}
\includegraphics[width=0.5\textwidth]{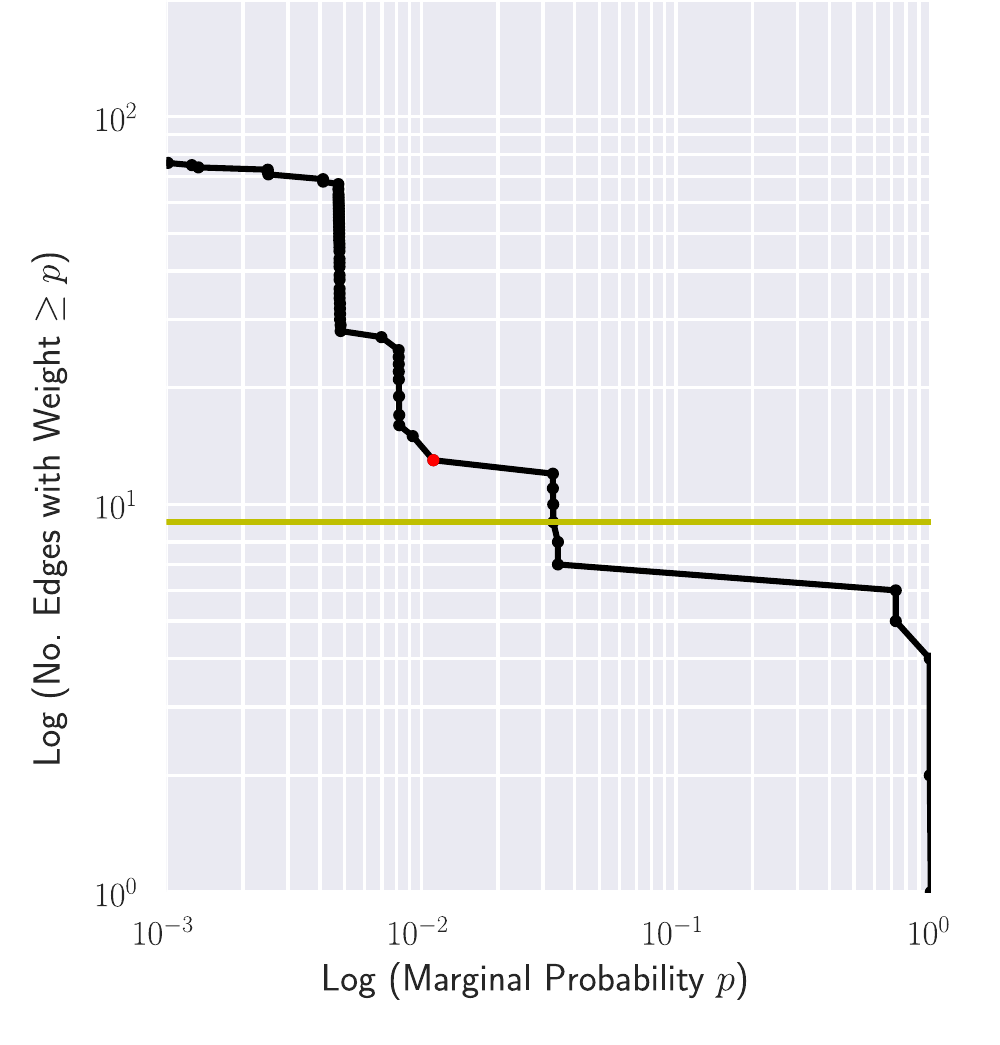}
\caption{\textbf{The distribution of the number of edges with weight $ \ge p$ in GBM with $k=3$ and $t=3$ in log-log scale.} The red dot indicates the first hitting edge weight where the change in slope is negative (when moving leftward) such that the number of edges in the subgraph is at least $t \times {k \choose 2} = 9$ (as the horizontal yellow line). }
\label{fig:delta_selection}
\end{center}
\end{figure*}

\begin{figure*}[h!]
\begin{center}
\includegraphics[width=\textwidth]{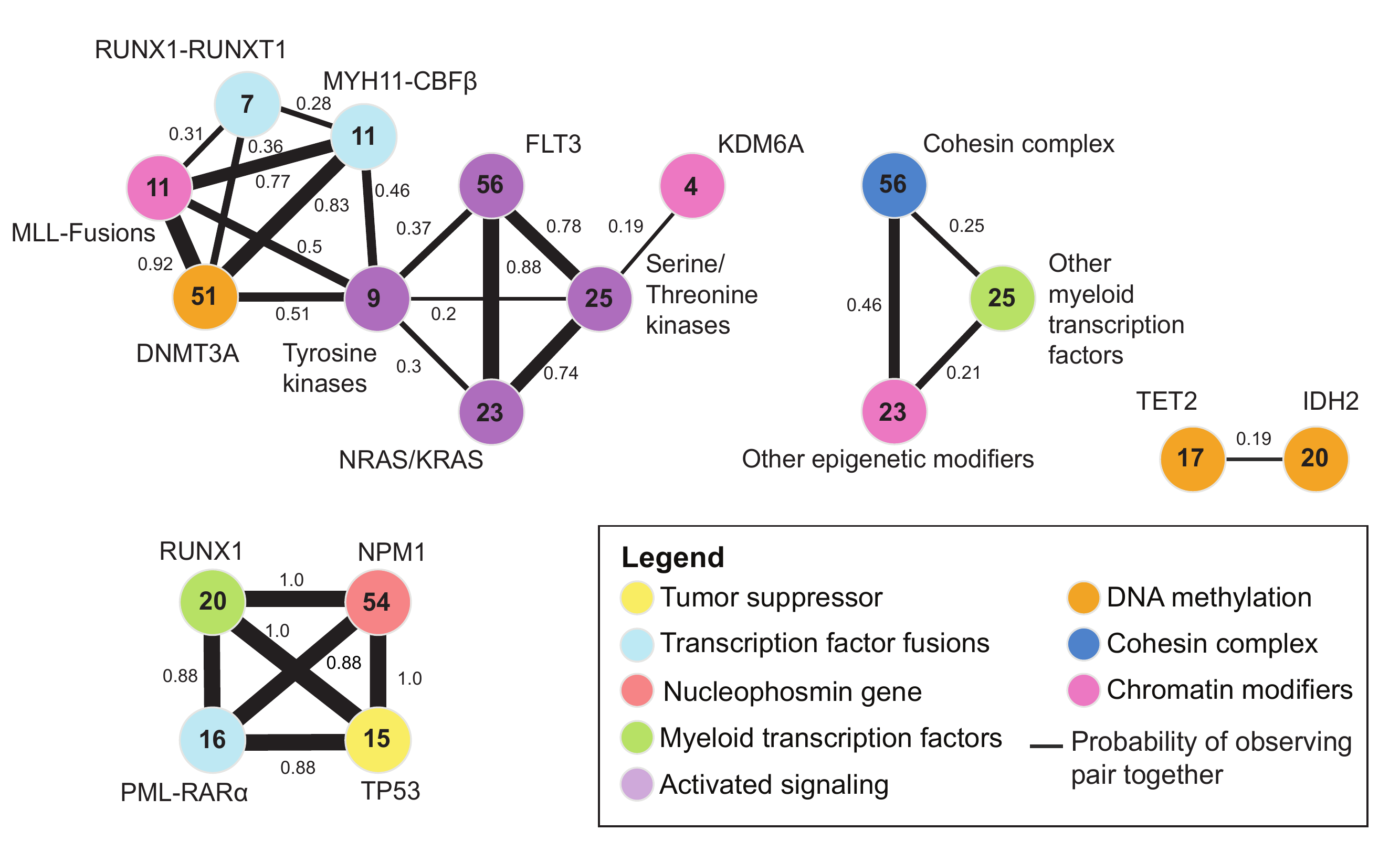}
\caption{\textbf{\OurAlgo{} results on AML dataset with $t=4$ and $k=4$. }}
\label{fig:aml-results}
\end{center}
\end{figure*}

\begin{figure*}[h!]
\begin{center}
\includegraphics[width=\textwidth]{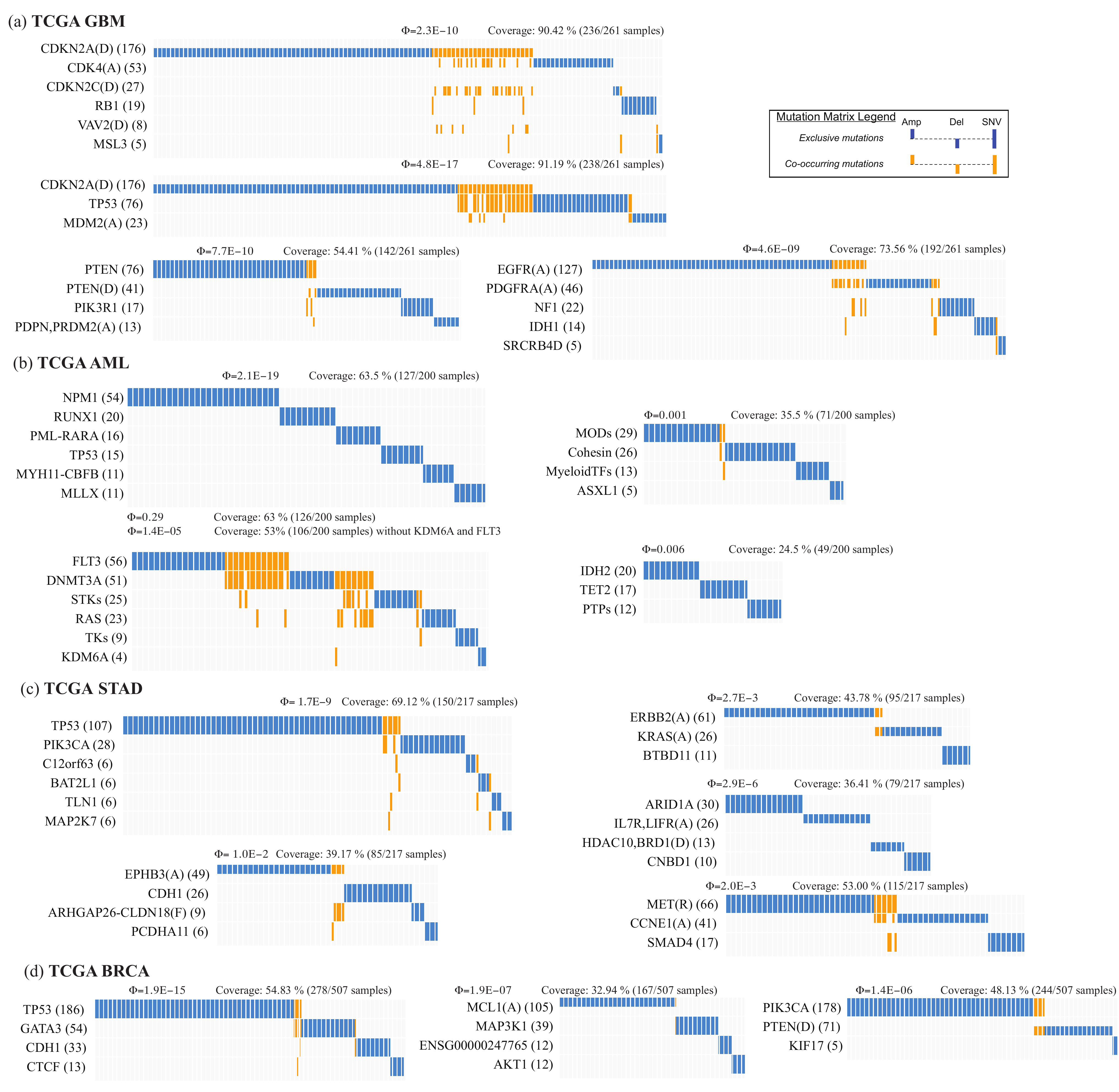}
\caption{\textbf{Mutation matrices for the \OurAlgo{} results on (a) TCGA GBM, (b) TCGA AML, (c) TCGA STAD, and (d) TCGA BRCA datasets.} The matrices have alterations as rows, and samples as columns. Each cell indicates whether or not an alteration occurred in a particular sample, where grey indicates the sample was not altered. Samples with co-occurring alterations in the same set are colored orange, while exclusive alterations are colored blue.}
\label{fig:mutation-matrices}
\end{center}
\end{figure*}

 \end{document}